\documentclass[aps,pre,showpacs,twocolumn,preprintnumbers,amsmath,amsfonts,bm]{revtex4-1}
\usepackage{color}
\usepackage{graphicx}
\usepackage{subfigure}
\usepackage{dcolumn}
\usepackage{color}
\usepackage{tikz}
\usepackage{amsmath}
\usetikzlibrary{arrows,automata,backgrounds,calendar,chains,matrix,mindmap,patterns,petri,shadows,shapes.geometric,shapes.misc,trees,plotmarks}
\graphicspath{{./plots/}}

\begin{document}

\title{Computational Capabilities of Random Automata Networks for Reservoir Computing}

\author{David Snyder$^{1}$, Alireza Goudarzi$^{2}$, Christof Teuscher$^{1}$}

\affiliation{
$^1$Portland State University, 1900 SW $4^{th}$ Avenue, Portland, OR 97206 USA\\
$^2$University of New Mexico, 1 University Boulevard Northeast Albuquerque, NM 87131 USA\\
}

\date{\today}

\begin{abstract}
This paper underscores the conjecture that intrinsic computation is maximal in systems at the ``edge of chaos." We study the relationship between dynamics and computational capability in Random Boolean Networks (RBN) for Reservoir Computing (RC). RC is a computational paradigm in which a trained readout layer interprets the dynamics of an excitable component (called the reservoir) that is perturbed by external input. The reservoir is often implemented as a homogeneous recurrent neural network, but there has been little investigation into the properties of reservoirs that are discrete and heterogeneous. Random Boolean networks  are generic and heterogeneous dynamical systems and here we use them as the reservoir. An RBN is typically a closed system; to use it as a reservoir we extend it with an input layer. As a consequence of perturbation, the RBN does not necessarily fall into an attractor. Computational capability in RC arises from a trade-off between separability and fading memory of inputs. We find the balance of these properties predictive of classification power and optimal at critical connectivity. These results are relevant to the construction of devices which exploit the intrinsic dynamics of complex heterogeneous systems, such as biomolecular substrates.

\end{abstract}


\maketitle

\section{Introduction}
Reservoir computing is an emerging paradigm that promotes computing
using the intrinsic dynamics of an excitable system called the
reservoir \citep{Lukosevicius:2009p1443}.  The reservoir acts as a
temporal kernel function, projecting the input stream into a higher
dimensional space, thereby creating features for the readout layer.
To produce the desired output, the readout layer performs a
dimensionality reduction on the traces of the input signal in the
reservoir. Two advantages of RC are: computationally inexpensive
training and flexibility in reservoir
implementation. {The latter is particularly important
  for systems that cannot be designed in a top-down way by traditional
  engineering methods. RC permits computation with physical systems that
  show extreme variation, interact in partially or entirely unknown
  ways, allow for limited functional control, and have a dynamic
  behavior beyond simple switching.} This makes RC suitable for
emerging unconventional computing paradigms, such as computing with
physical phenomena \citep{Fernando2003} and self-assembled electronic
architectures \citep{Haselman:2010p657}. {The
  technological promise of harnessing intrinsic computation with RC
  beyond the digital realm has enormous potential for cheaper, faster,
  more robust, and more energy-efficient information processing
  technology.}

 \citet{Maass:2002p1444} initially proposed
a version of RC called Liquid State Machine (LSM) as a model of
cortical microcircuits.  Independently, \citet{Jaeger:2001p1442}
introduced a variation of RC called Echo State Network (ESN) as an
alternative recurrent neural network approach for control tasks.
Variations of both LSM and ESN have been proposed for many different
machine learning and system control tasks
(\citet{Lukosevicius:2009p1443}). Insofar, most of the RC research is
focused on reservoirs with homogeneous in-degrees and transfer
functions. However, due to high design variation and the lack of
control over these devices, most self-assembled systems are
heterogeneous in their connectivity and transfer functions.

Since RC can be used to harness the intrinsic computational
capabilities of physical systems, our study is motivated by three
fundamental questions about heterogeneous reservoirs:
\begin{enumerate} 
\item What is the relationship between the dynamical properties of a
  heterogeneous system and its computational capability as a
  reservoir?
\item How much does a reservoir need to be perturbed to adequately
  distribute the input signal? It may be infeasible to perturb the
  entire system. Also, a single-point perturbation may not propagate
  throughout the system due to its internal topology.  Thus, we
  consider the size of the perturbation necessary to adequately
  distribute the input signal.
\item In a physical RC device, it may be difficult to observe the
  entire system. How much of the system and which components ought to
  be observed to extract features about the input stream?
\end{enumerate}

We model the reservoirs with Random Boolean Networks (RBN), which are
chosen due to their heterogeneity, simplicity, and generality.
\citet{Kauffman:1969p2786} first introduced this model to study gene
regulatory networks. He showed these Boolean networks to be in a
complex dynamical phase at ``the edge of chaos" when the average
connectivity (in-degree) of the network is $\langle K\rangle=2$
(critical connectivity).  \citet{rohlf07_prl} showed that with
near-critical connectivity information propagation in Boolean networks
becomes independent of system size. \citet{packard1988} used an
evolutionary algorithm to evolve Cellular Automata (CA) for solving
computational tasks. He found the first evidence that connects
critical dynamics and optimal computation in CA.  Detailed analysis by
\citet{mitchell93} refuted this idea and accounted genetic drift, not
the CA dynamics, for the evolutionary behavior of the CA.
\citet{PhysRevLett.108.128702} studied adaptive computation and task
solving in Boolean networks and found that learning drives the network
to the critical connectivity $\langle K_{c}\rangle=2$.

\citet{snyder2012} introduced RBNs for RC, and found optimal task
solving in networks with $\langle K \rangle > \langle K_{c} \rangle$.
Here, using a less restrictive RC architecture, we find that RBNs with
critical dynamics provided by $\langle K_{c} \rangle$ tend to offer
higher computational capability than those with ordered or chaotic
dynamics.

To be suitable for computation, a reservoir needs to eventually forget past 
perturbations, while possessing dynamics which
respond in different ways due to different input streams. The first requirement is 
captured by \textit{fading memory}. The \textit{separation property} captures the second requirement and computes a distance measurement between the states of two identical reservoirs after being perturbed by two distinct input streams. It has been hypothesized that computational capabilities are optimal when the \textit{separation property} is highest, but old input is eventually forgotten by the reservoir, which occurs when \textit{fading memory} is lowest \cite{natschlager2004}. We extend the measurements described in \cite{natschlager2004,4905041020100501} to predict the computational capability of a reservoir in finite time-scales with a short-term memory requirement.

\section{Model}
A Reservoir Computing device is made up of three parts: input layer, reservoir, 
and readout layer [cf. Fig.~\ref{fig:rc}].  The input layer
excites the reservoir by passing an input signal to it, and the readout
layer interprets the traces of the input signal in the reservoir 
dynamics to compute the desired output.  In our model, the reservoir
is a Random Boolean Network (RBN).  The fundamental subunit of an RBN is a 
node with $K$ input connections.  At any instant in time, the node can assume 
either of the two binary states, ``0'' or ``1.'' The node updates its state at time $t$ 
according to a $K$-to-1 Boolean mapping of its $K$ inputs.  Therefore, the state 
of a single node at time $t+1$ is completely determined by its $K$ inputs at time 
$t$ and by  one of the
$2^{2^{K}}$ Boolean functions used by the node. An RBN is a collection of $N$ 
such binary nodes.  For each node $i$ out of $N$ nodes, the node receives 
$K_{i}$ inputs, each of which is connected to one of the $N$ nodes in the 
network.  In this model, self-connections are allowed.

The network is random in two different ways: 1) the source nodes for an
input are chosen from the $N$ nodes in the network with uniform
probability and 2) the Boolean function of node $i$ is chosen from the
$2^{2^{K_i}}$ possibilities with uniform probability. Each node sends
the same value on all of its output connections to the destination
nodes. The average connectivity will be $\langle
K\rangle=\frac{1}{N}\sum_{i=1}^N{K_i}$. We study the properties of RBNs
characterized by $N$ nodes and average connectivity $\langle
K\rangle$. This refers to all the instantiations of such RBNs. Once
the network is instantiated, the collective time evolution at time $t$
can be described as using
$x_i^{t+1}=f_i(x_1^t,x_2^t,\dots,x_{K_i}^t)$, where $x_i^t$ is the
state of the node $i$ at time $t$ and $f_i$ is the Boolean function
that governs the state update of the node $i$. The nodes are updated
synchronously, i.e., all the nodes update their state according to a
single global clock signal.

{From a graph-theoretical perspective, an RBN is a directed graph with
$N$ vertices and $E=\langle K\rangle N$ directed edges. 
  We construct the graph according to the random graph model \protect
\citep{Erdos:1959p1849}. We call this model a heterogeneous RBN because 
each node has a different in-degree. In the classical RBN model, all the nodes 
have identical in-degrees and therefore are homogeneous. The original model of 
{\citet{Kauffman:1969p2786} assumes a static environment and therefore does 
not include exogenous inputs to the network. To use RBNs as the reservoir, we 
introduced $I$ additional input nodes that each distribute the input signals to $L$ 
randomly picked nodes in the network. The source nodes of $K_i$ links for each 
node $i$ are randomly picked from $N$ nodes with uniform probability.} The 
input nodes are not counted in calculating $\langle K\rangle$. {For online
computation, the reservoir is extended by a separate readout layer
with $O$ nodes. Each node in the readout layer is connected to each node in the 
reservoir. The output of node $o$ in the readout layer at time $t$ is denoted by 
$y_o^t$ and is computed according to $y_o^t=sign\left(\sum_{j=1}
^N{\alpha_jx_j^t}+b\right)$. Parameters $\alpha_j$ are the weights on the inputs 
from node $j$ in the reservoir to node $o$ in the readout layer, and $b$ is the 
common bias for all the readout nodes. Parameters $\alpha_j$ and $b$ can be
trained using any regression algorithm to compute a target output
\citep{Jaeger:2001p1442}. In this paper, we are concerned with RBN-RC devices 
with a single input node, and a single output node.

 \begin{figure}
 \includegraphics[]{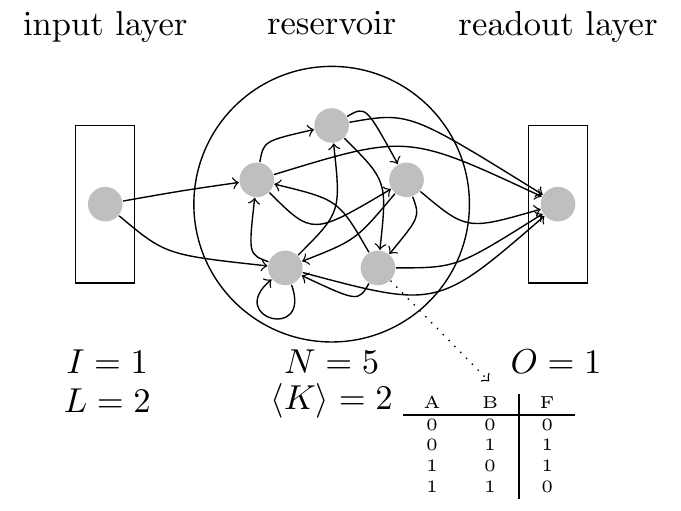}

  \caption{Schematic of a reservoir computing system. The input layer
 delivers the input signals to random nodes inside the reservoir. The
 readout layer receives output signals from random nodes inside the
 reservoir.
  The reservoir itself is made of a collection of computing
 nodes that are randomly interconnected. The reservoir creates a
 representation of the input signals that can be read and classified by
 the readout layer. Learning is performed by training only the readout
 layer nodes and connections.
 }
  \label{fig:rc}
  \end{figure}

\section{Measures}
\subsection{Perturbation Spreading}

RBNs are typically studied as closed systems in which the notion of damage 
spreading is used to classify the RBNs' dynamics as ordered, critical, or chaotic 
\citep{Derrida:1986}. Because our model requires external perturbations, we 
must extend the notion of damage spreading to account for RBNs which are 
continuously excited by external input. Since an RBN used as a reservoir is
not a closed system, the propagation of external perturbations may behave 
distinctly from the propagation of damage in the initial states of the RBN. Let $
\mathcal{M}$ be an RBN with $N$ nodes and average connectivity $\langle K 
\rangle$.  Let $u_{a}$ be an input stream, and $u_{b}$ be a variation of $u_{a}$.  
Then the perturbance spreading of $\mathcal{M}$ with an input stream $u_{a}$ 
and its variation $u_b$ is $\mathcal{P}(\mathcal{M}, u_{a}, u_{b}) = \frac{H(A,B)}
{N}$, where $A$ and $B$ are the states of the RBN after being driven by input 
streams $u_{a}$ and $u_{b}$ respectively, and $H(A,B)$ is the Hamming 
distance between the states.

For a dynamical system to act as a reservoir, it needs to be excited in different ways by very different input streams, while
eventually forgetting past perturbations.  These measurements are captured by the notions of \textit{separation} and \textit{fading memory} in \citep{natschlager2004}. However, to account for the importance of short-term memory in the reservoir and finite-length input streams, we are specifically interested in the \textit{separation} of the system $\tau$ time steps in the past, within an input stream of length $\mathcal{T}$.

The ability of the RBN to separate two input streams of length $\mathcal{T}$, which differ for only the first $\mathcal{T} - \tau$ time steps, is given by 
\begin{equation}
\mathcal{S}_{\tau}(\mathcal{M},\mathcal{T}) = \mathcal{P}(\mathcal{M}, u, v),
\end{equation} where $\mathcal{T} = |u| = |v|$ and $$v_{i} =
\begin{cases}
\bar{u}_{i}, & \text{if } i < \mathcal{T} - \tau \\
u_{i}, & \text{otherwise.}
\end{cases}
$$

In order for an RC device to be able to generalize, a reservoir needs to eventually forget past perturbations.  Thus we define:

\begin{equation}
\mathcal{F}(\mathcal{M}, \mathcal{T}) = \mathcal{P}(\mathcal{M}, u, w),
\end{equation} 

where $\mathcal{T} = |u| = |w|$ and $$w_{i} =
\begin{cases}
\bar{u}_{i}, & \text{if } i = 0 \\
u_{i}, & \text{otherwise.}
\end{cases}
$$

\citet{natschlager2004} found that computational capability of recurrent neural network reservoirs are greatest when the difference between 
\textit{separation} and \textit{fading memory} are largest and that this coincides with critical dynamics. Therefore, we want fading memory to
be low, while separation is high.  We define the computational capability of a reservoir $\mathcal{M}$ over an input stream of length $\mathcal{T}$, $\tau$ time steps in the past as:
\begin{equation}
\Delta(\mathcal{M}, \mathcal{T}, \tau) = \mathcal{S}_{\tau}(\mathcal{M},\mathcal{T}) - \mathcal{F}(\mathcal{M},\mathcal{T}).
\end{equation}

\subsection{Entropy and Mutual Information}
Information theory \cite{Shannon:1948} provides a generic framework for measuring information transfer, noise, and loss between a source and a destination. The fundamental quantity in information theory is Shannon information defined as the entropy $\mathcal{H}_S$ of an information source $S$. For a source $S$ that takes a state $\{s_i|1\le i\le n\}$ with probability $p(s_i)$, the entropy is defined as:
\begin{equation}
\mathcal{H}_S=-\sum_{i=1}^{n}{p(s_i)\log_2p(s_i)}.
\label{eq:entropy}
\end{equation}  
This is the amount of information that $S$ contains. To measure how much information is transferred between a source and a destination, we calculate the mutual information $\mathcal{I}(S:D)$ between a source $S$ and a destination $D$ with states $d_j$. Before we can calculate $\mathcal{I}(S:D)$ we need to calculate a joint entropy of the source and destination as follows:
\begin{equation}
\mathcal{H}_{SD}=-\sum_{i=1}^n{\sum_{j=1}^m{p(s_i,d_j)}\log_2p(s_i,d_j)}.
\end{equation}
Now the mutual information is given by:
\begin{equation}
\mathcal{I}(S:D)=\mathcal{H}_S+\mathcal{H}_D-\mathcal{H}_{SD}.
\end{equation}
We will see later how we can use entropy and mutual information to see how much information from the input signals are transferred to the reservoir and how much information the reservoir can provide about the output while it is performing computation.

\subsection{Tasks}

We use the temporal parity and density classification tasks 
to test the performance of the reservoir systems. According to the task, the RC
system is trained to continuously evaluate $n$ bits which were
injected into the reservoir beginning at $\tau + n$ time steps in the
past.

\subsubsection{Temporal Parity} The task determines if $n$ bits $\tau + n$
to $\tau$ time steps in the past have an odd number of ``1'' values. Given an input stream $u$, where $|u| = \mathcal{T}$, a delay $\tau$, and a window $n \ge 1$,

\begin{displaymath} \mathcal{A}_{n}(t) = \left\{
     \begin{array}{ll} u(t - \tau), & \text{if } n = 1\\
\displaystyle\oplus_{i=0}^{n-1}{u(t - \tau - i)}, & \text{ otherwise,}
     \end{array} \right.
\end{displaymath}
where $ \tau + n \leq t \leq \mathcal{T} - \tau - n$.

\subsubsection{Temporal Density} The task determines whether or not an
odd number of bits $\tau + n$ to $\tau$ time steps in the past have
more ``1" values than ``0." Given an input stream $u$, where $|u| =
\mathcal{T}$, a delay $\tau$, and a window $n = 2k + 1$, where $k \ge 1$,

 \begin{displaymath}
   \mathcal{B}_{n}(t) = \left\{
     \begin{array}{ll}
       1, & \text{if }  2\displaystyle\sum_{i=0}^{n-1}{u(t - \tau - i)} > n \\
       0, & \text{ otherwise,}
     \end{array}
   \right.
\end{displaymath} 
where $ \tau + n \leq t \leq \mathcal{T} - \tau - n$.

\subsubsection{Training and Evaluation} For every system, we randomly generate a training set $S_{T}$
and testing set $S_{G}$.  For each stream $v \in S_{T}$ or $u \in S_{G}$, $|v| = |u| = \mathcal{T}$. The size of the training and testing sets are dependent on $n$, and determined by the following table.

\begin{center}
    \begin{tabular}{ | p{1cm} | p{2cm} | p{2cm} |}

    \hline
    $n$ & $|S_{T}|$ & $|S_{G}|$ \\ \hline
    1 & 50 & 50   \\ \hline
    3 & 150 & 150   \\ \hline    
    5 & 300 & 150    \\ \hline
    7 & 500 & 150    \\ \hline
    9 & 500 & 150    \\ \hline
    \end{tabular}
\end{center}

We train the output node with a form of stochastic gradient descent in which
the weights of the incoming connections are adjusted after every time
step in each training example.  Given our system and tasks, this form
of gradient descent appears to yield better training and testing
accuracies than the conventional forms.  We use a learning rate $\eta
= 0.01$, and train the weights for up to 20,000 epochs. Since the dynamics of the 
underlying RBN are deterministic and reset after each training stream,
we terminate training early if an accuracy of $1.0$ is achieved on $S_{T}$.  The 
accuracy of an RC device on a stream $v \in S_{T}$ is determined by the number 
of times that the output matches the expected output as specified by the task 
divided by the total number of values in the output stream.  The accuracy on each
input set is summed together and divided by the total number of input streams in 
the set to calculate the current training accuracy $T$. After the weights of the 
output layer are trained on the input streams in $S_{T}$, they remain fixed. We 
then drive the reservoirs with input streams $u \in S_{G}$ and record the number 
of times that the output of the RC device matches the expected output. The 
generalization capability $G$ is then computed by dividing the total number of 
times in which the output of the readout layer matches the correct output, by the 
total number of correct outputs. This process is averaged over all streams in 
$S_{G}$. In general, we are interested in finding the reservoirs that maximize $G
$.

\section{Results}
\subsection{Computational Capability}

\begin{figure}[t]
\centering
\subfigure[]{
\includegraphics[width=1.55in]{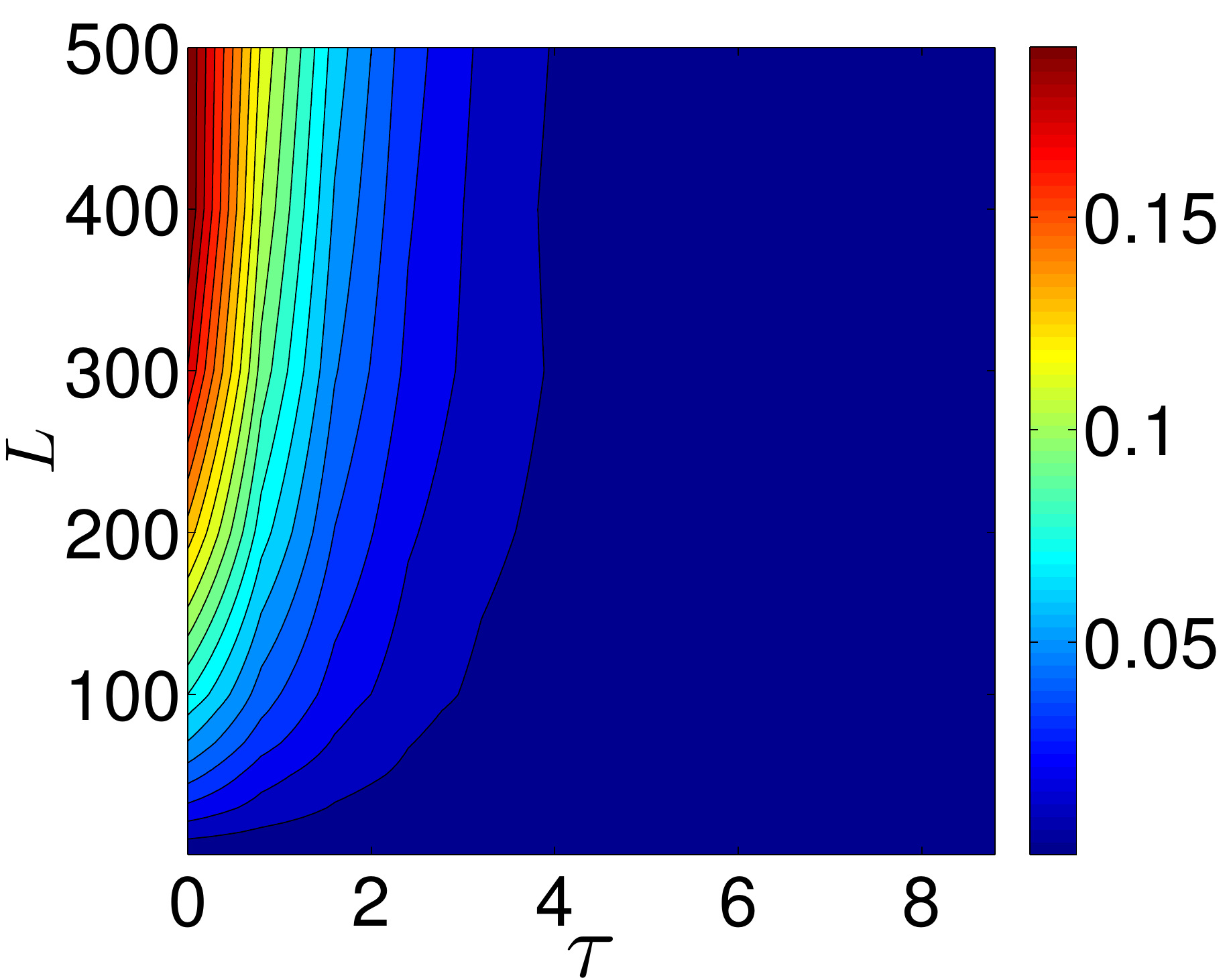}
\label{fig:K1dot0T103D}
}
\subfigure[]{
\includegraphics[width=1.55in]{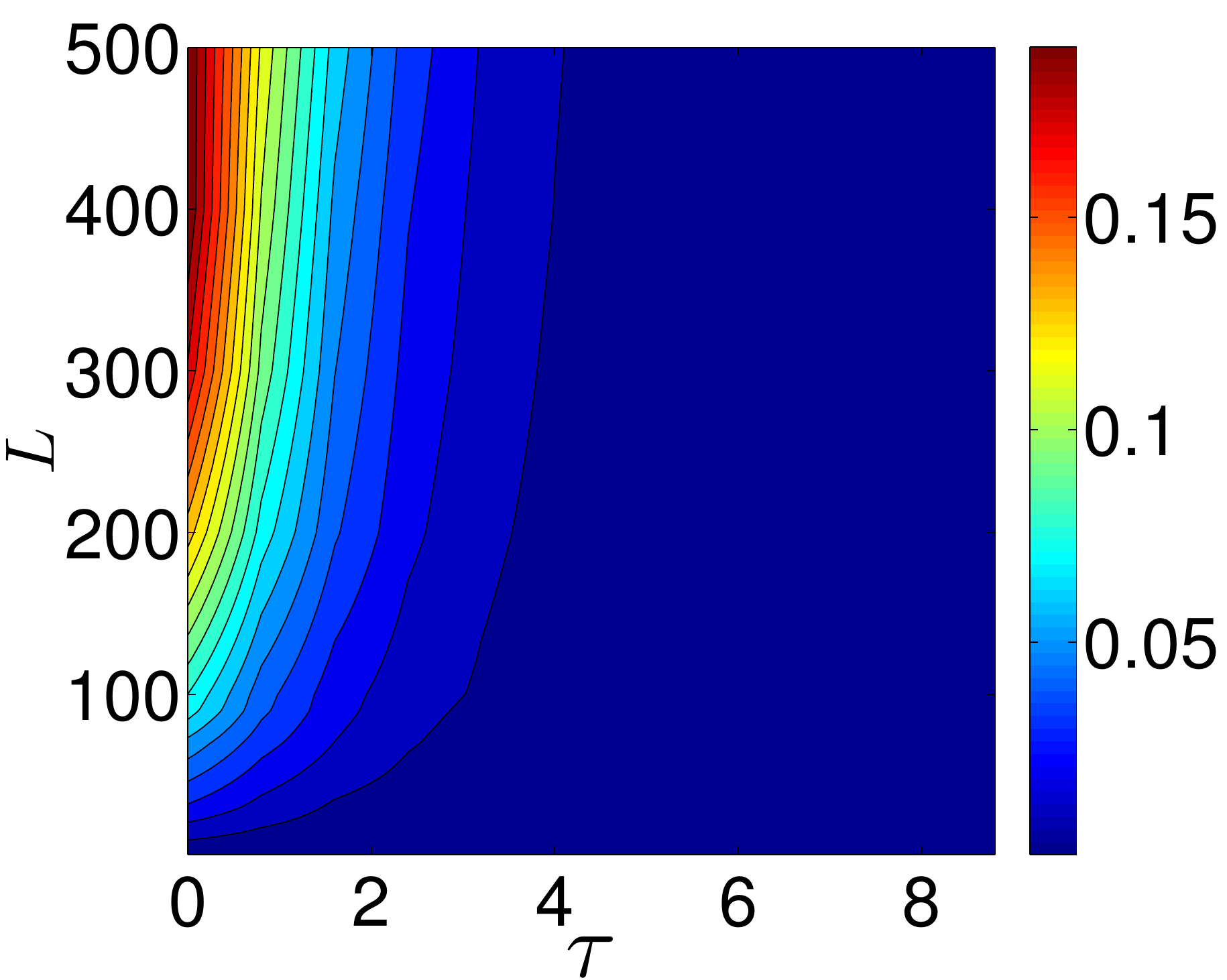}
\label{fig:K1dot0T1003D}
}
\subfigure[]{
\includegraphics[width=1.55in]{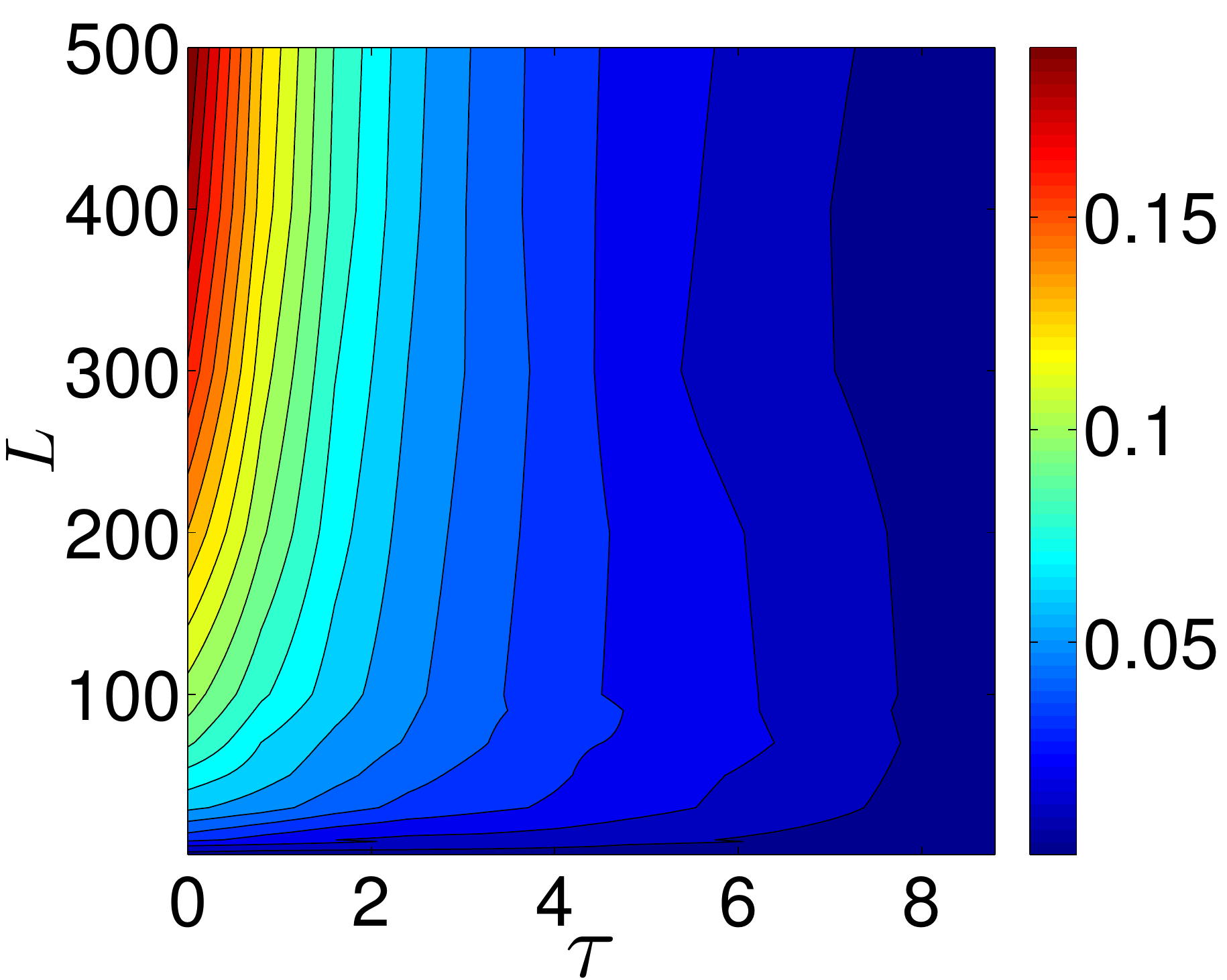}
\label{fig:K2dot0T103D}
}
\subfigure[]{
\includegraphics[width=1.55in]{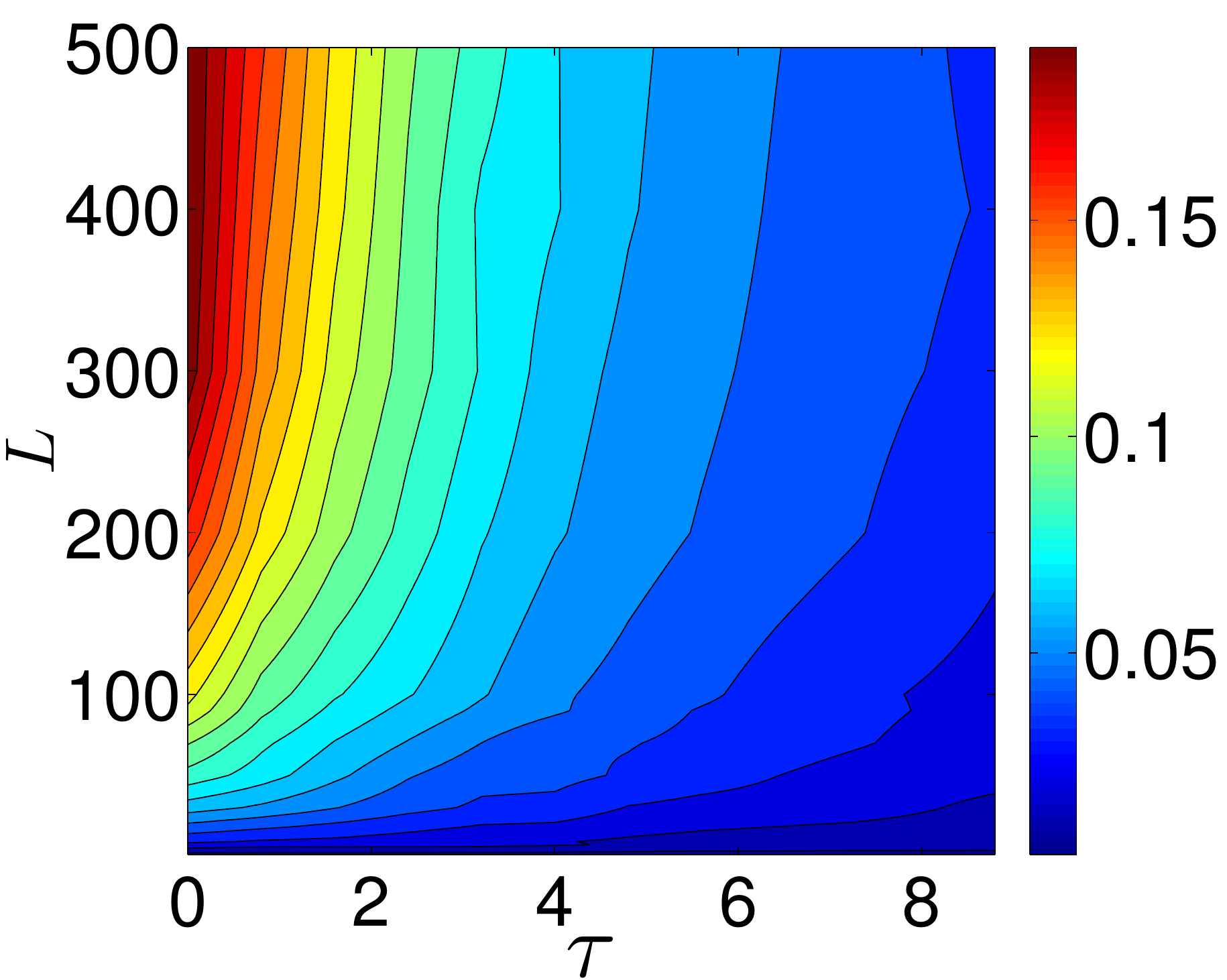}
\label{fig:K2dot0T1003D}
}
\subfigure[]{
\includegraphics[width=1.55in]{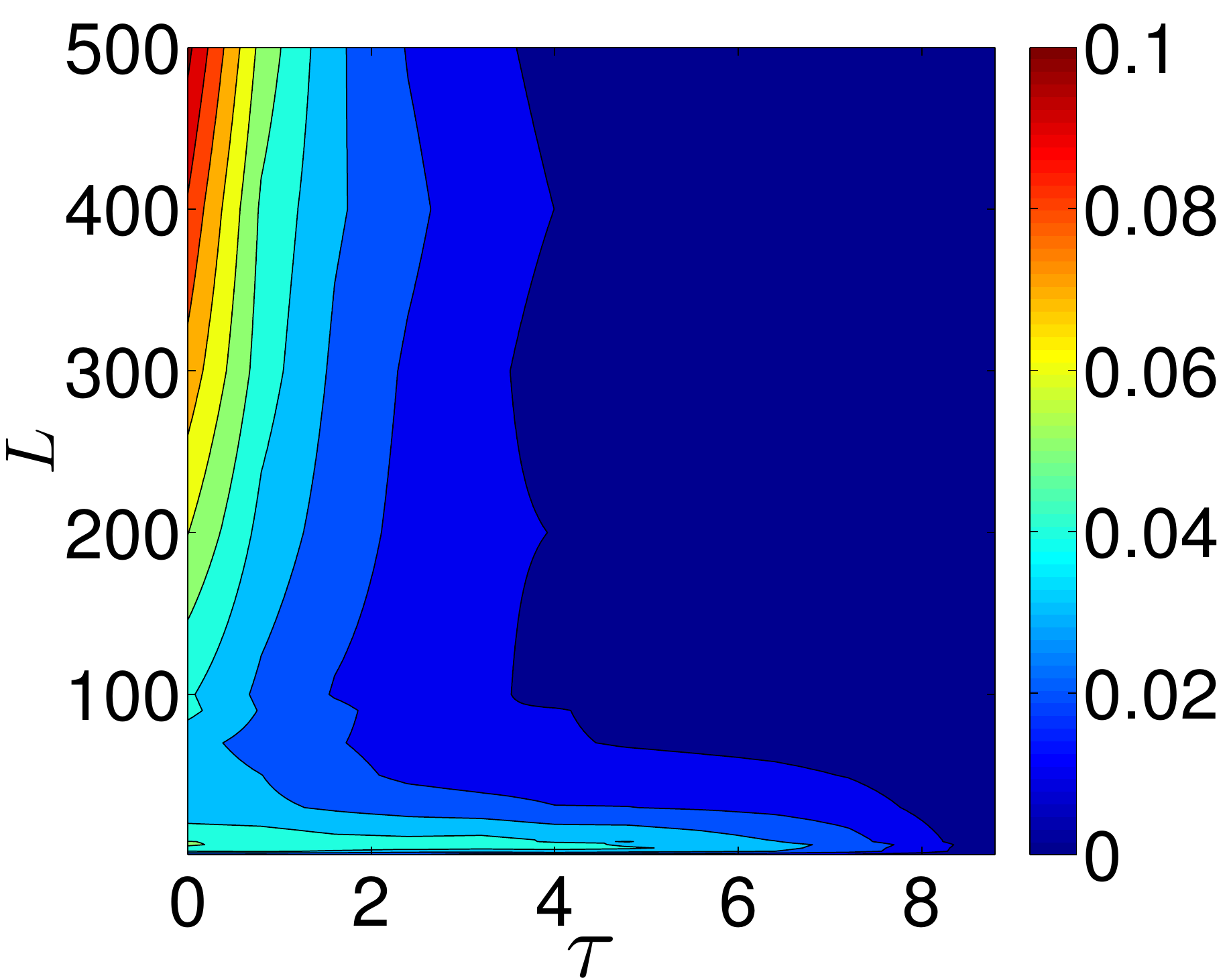}
\label{fig:K3dot0T103D}
}
\subfigure[]{
\includegraphics[width=1.55in]{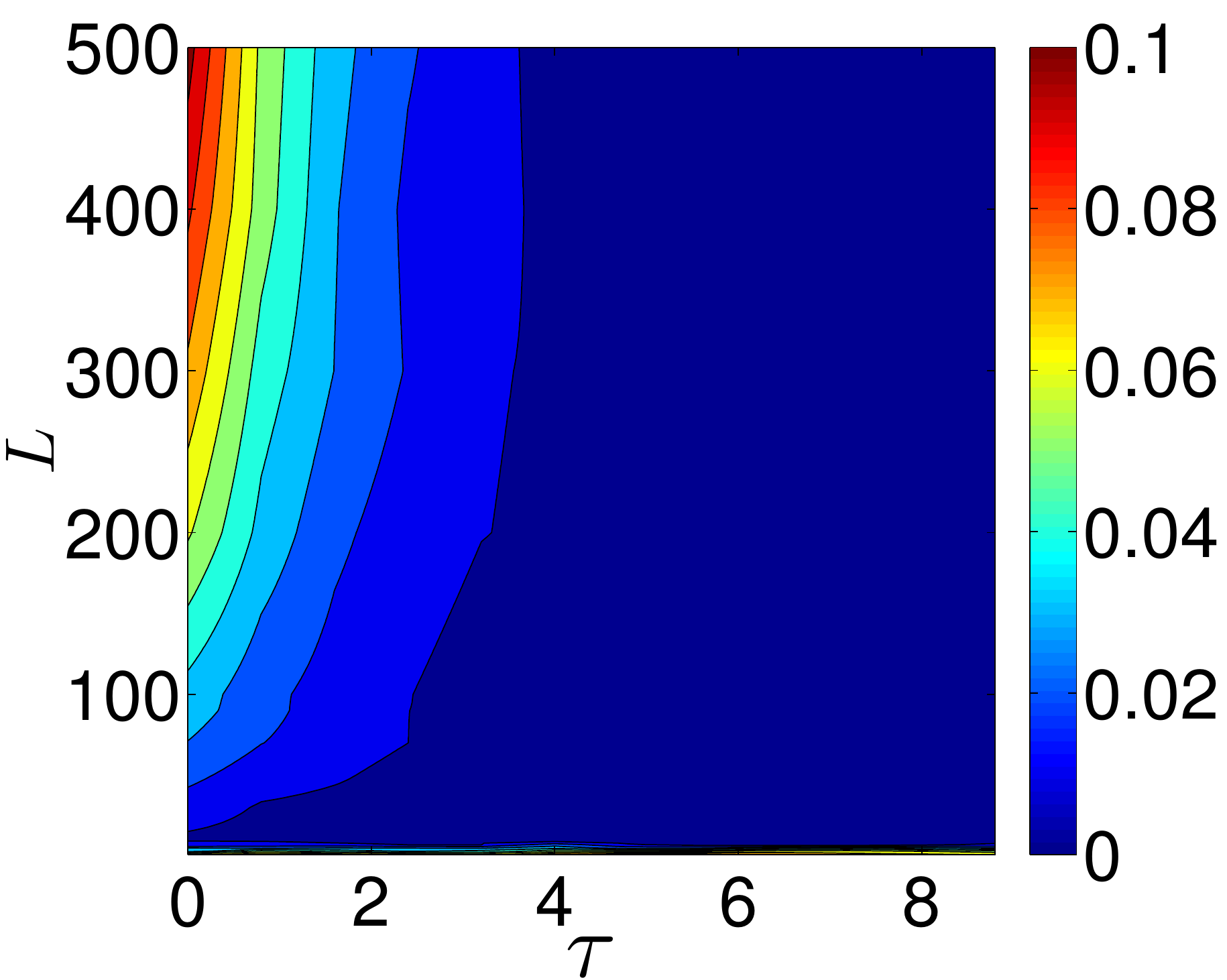}
\label{fig:K3dot0T1003D}
}

\caption{(Color online) The computational capability $\Delta$ of RBN reservoirs 
with $N=500$, $L \in [1,N]$ and $\tau \in [1,9]$. Parameters $\langle K \rangle$ 
and $\mathcal{T}$ are: $\langle K \rangle = 1$, $\mathcal{T} = 10$~
\subref{fig:K1dot0T103D}, $\langle K \rangle = 1$, $\mathcal{T} = 100$~
\subref{fig:K1dot0T1003D}, $\langle K \rangle = 2$, $\mathcal{T} = 10$~
\subref{fig:K2dot0T103D}, $\langle K \rangle = 2$, $\mathcal{T} = 100$~
\subref{fig:K2dot0T1003D}, $\langle K \rangle = 3$, $\mathcal{T} = 10$~
\subref{fig:K3dot0T103D}, $\langle K \rangle = 3$, $\mathcal{T} = 100$~
\subref{fig:K3dot0T1003D}. The computational capability $\Delta$ varies 
according to $\langle K \rangle \in \{1, 2, 3 \}$ and $\mathcal{T} = 10$
in the \textit{left} column and $\mathcal{T} = 100$ in the \textit{right} column.}
\label{fig:delta_L_K_tau}
\end{figure}

\begin{figure}[t]
\centering
\subfigure[]{
\includegraphics[width=1.55in]{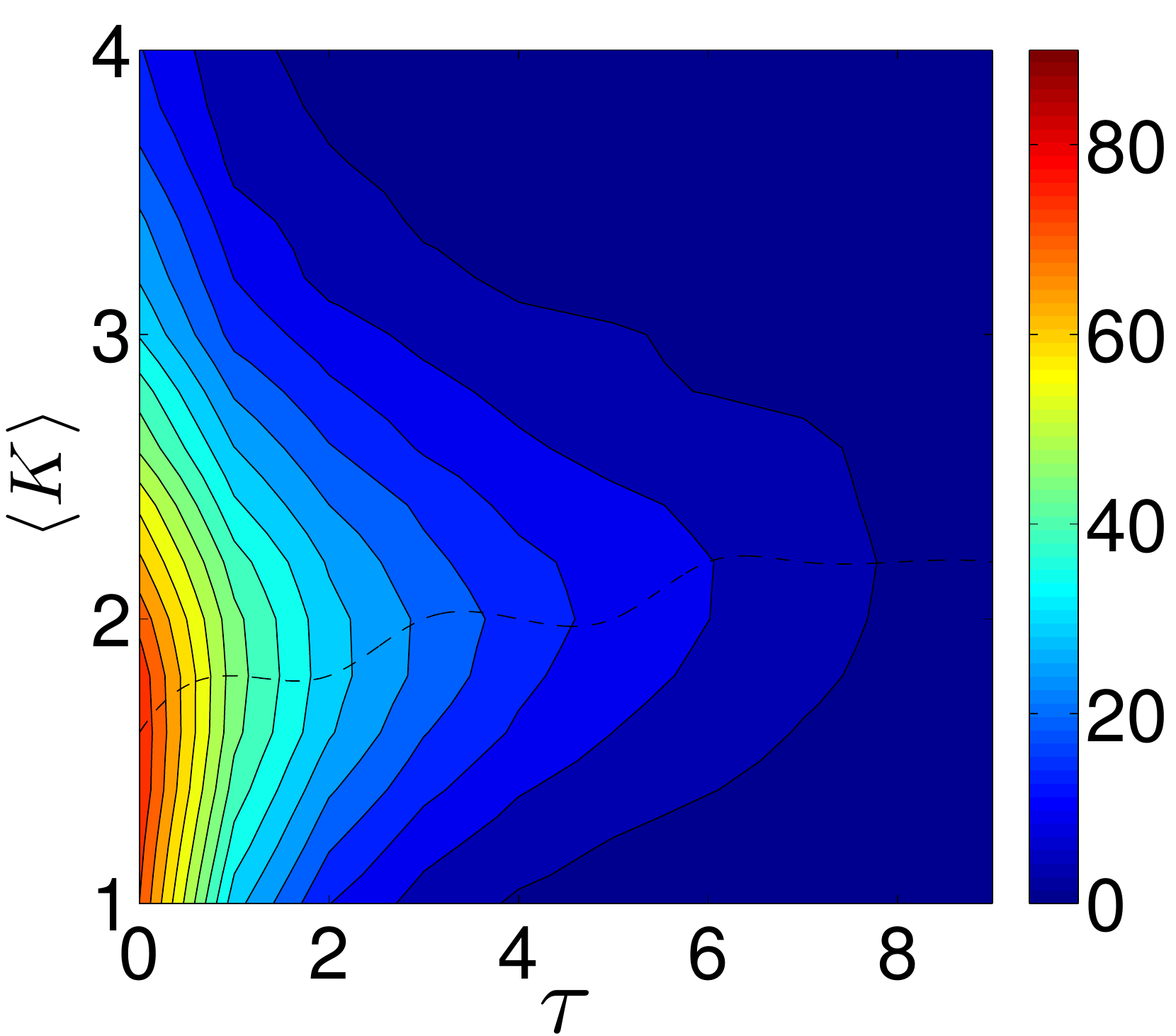}
\label{fig:KsT103D}
}
\subfigure[]{
\includegraphics[width=1.55in]{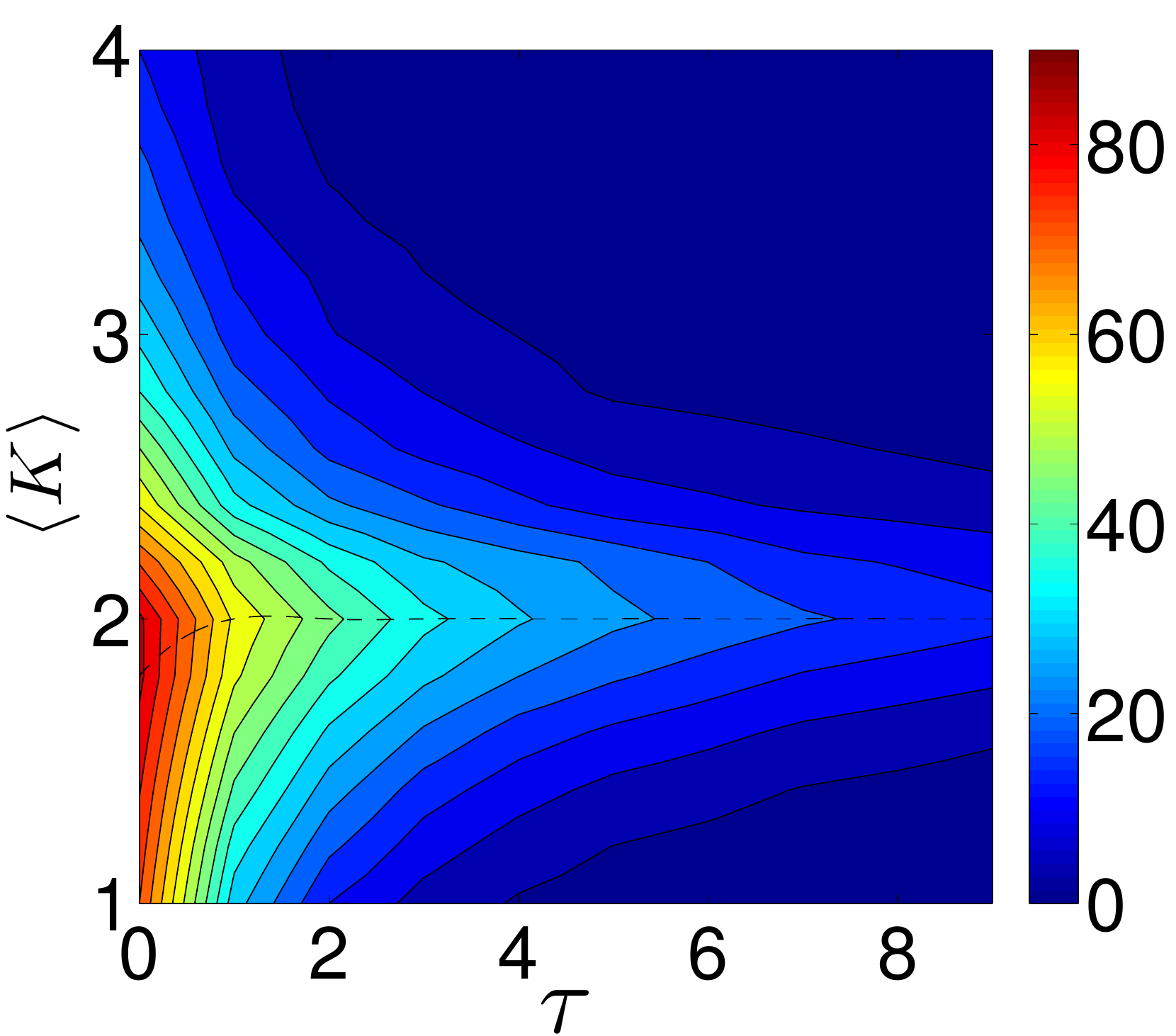}
\label{fig:KsT1003D}
}
\caption{(Color online) The computational capability $\Delta$ of RBN reservoirs $\mathcal{M}$ with $N=500$
and $\langle K \rangle \in [1, 4]$ summed over $L \in [1, 500]$. These are calculated as: $\sum_{L=1}^{N}{\Delta( {\mathcal{M}}, {\mathcal{T}}, \tau )}$, where $\mathcal{T} = 10$~\subref{fig:KsT103D}, and $\sum_{L=1}^{N}{\Delta( {\mathcal{M}}, {\mathcal{T}}, \tau )}$, where $\mathcal{T} = 100$~\subref{fig:KsT1003D}.
The dashed curve is a spline fit to the highest $~\Delta$, illustrating that near-critical 
connectivity maximizes computational capability, particularly in high $\mathcal{T}$.}
\label{fig:delta_K_tau}
\end{figure}

The computational capability as predicted by $\Delta$ are dependent on the 
properties of the reservoir $\mathcal{M}$, the length of the input stream $
\mathcal{T}$, and the memory $\tau$ required by the reservoir. The properties of 
$\mathcal{M}$ are determined by the dynamics which are due primarily to $
\langle K \rangle$ and the number of nodes $L$ which the input directly perturbs. 
For each $L$, $\langle K \rangle$, $\mathcal{T}$, and $\tau$ we calculate the 
average $\Delta(\mathcal{M}, \mathcal{T}, \tau)$ of $50$ instantiations of $
\mathcal{M}$. In figure~\ref{fig:delta_L_K_tau} we present these results for $
\langle K \rangle \in \{1,2,3\}$. To produce Fig.~\ref{fig:delta_K_tau} we sum 
over $\Delta(\mathcal{M}, \mathcal{T}, \tau)$ for all $L$. The dashed curves in 
Figs.~\ref{fig:KsT103D} and~\ref{fig:KsT1003D} are spline fits which highlight 
the greatest $\Delta$ values. In Figs.~\ref{fig:delta_L_K_tau} and~
\ref{fig:delta_K_tau} we see that RBNs with critical connectivity $\langle K_{c} 
\rangle = 2$ tend to provide the highest $\Delta$.  A high $\Delta$ signifies that 
the reservoir dynamics have the ability to separate different input streams, while 
having dynamics which are determined more by recent input, than past input. In 
contrast, a low $\Delta$ signifies either or both of the following: i) the reservoir's 
dynamics are too frozen to separate different input streams effectively or ii) traces 
of early perturbations are never forgotten by the reservoir. The consequence of i) 
is the inability to compute difficult tasks, such as $\mathcal{A}_{n}$ or those which 
require long short-term memory. The consequence of ii) is great difficulty in 
generalizing; if past information which is irrelevant to computing the correct 
output in the readout layer dominates the dynamics of the reservoir, the output 
layer will be unable to classify the dynamics caused by the more relevant, recent 
input.

We see that $\langle K \rangle = 1$ has a very high $\Delta$ only when $\tau$ is 
small.  This is due to the brief short-memory afforded to an RBN with subcritical 
dynamics. Since a network with $\langle K \rangle = 1$ has little short-term 
memory, its computational capabilities are unaffected by an increase in $
\mathcal{T}$, as demonstrated in Figs.~\ref{fig:K1dot0T103D} and~
\ref{fig:K1dot0T1003D}: there is no memory at all of early perturbations.

Chaotic reservoirs, represented here by $\langle K \rangle = 3$, are 
characterized by their sensitivity to initial perturbations, and a high 
\textit{separation}. In two identical, chaotic systems, a single bit difference in their 
respective input streams will eventually become magnified until the two systems 
differ by the states of some $P$ nodes. If the initial perturbation is larger than $P
$, then the differences in the systems will diminish until reaching $P$. Because of 
this, a chaotic system could maximize its $\Delta$ in two different ways: i) 
compute over a sufficiently short input stream and ii) perturb enough of the 
system so that the recent input has a more significant effect on the dynamics than 
the past input. In i), the restriction of a brief input stream can be relaxed if the 
input stream perturbs as few nodes as possible, giving the system more time to 
propagate perturbations (cf. Fig.~\ref{fig:K3dot0T103D}). On the other hand ii) 
requires maximizing $L$ (cf. Fig.~\ref{fig:K3dot0T1003D}). However, even if 
distortion is staved off by slowing down the propagation of external perturbations, 
the system is ultimately fated to disorder.

\subsection{Information and Optimal Perturbation}
In the traditional implementations of reservoir computing, all the nodes in the 
reservoir are connected to the source of the input signal. 
Many task specific and generic measures of computation in reservoirs have been 
comprehensively studied in \cite{4905041020100501}. 
However, the relationship between the computational properties of the reservoir 
and the number of nodes which the input layer perturbs remains unexplored.
Here, we use information theory to characterize the computation in the reservoir 
as information transfer between the input and the reservoir and between the 
reservoir and the output.

In reservoir computing, the reservoir is a dynamical system and therefore has 
intrinsic entropy. The input is also time-varying and we can calculate its entropy. 
In order to reconstruct the desired function, the output layer has to pick up the 
traces of the input in the reservoir dynamics. This fact is reflected in the entropy 
change of the reservoir due to its input and therefore can be measured using 
mutual information between the input $I$ and the reservoir $R$, i.e., $\mathcal{I}
(I:R)$. In our study we distribute the input to the reservoir only sparsely, we would 
thus like to find how $\mathcal{I}(I:R)$ changes as a function of $L$ and if there is 
an optimal $L$. Moreover we would like to know, given a task to be solved, how 
much information the reservoir can provide to the output. That is, given the 
desired output, can the reservoir state be predictive of the output? This is 
equivalent to determining  how much information is transferred from the reservoir 
to the desired output. We show this measure using $\mathcal{I}(R:O)$ where $O$ 
indicates the output as the target.

In order to calculate $\mathcal{I}(I:R)$ and $\mathcal{I}(R:O)$, we consider the 
instantaneous states of the reservoir and its output to calculate the entropy. 
For the input, we need to calculate the entropy over the states that the input can 
take over the window size $n$. For example, on a input stream of length $
\mathcal{T}=50$ bits, window size $n$, and time delay $\tau=1$,  the input 
pattern $u_{t_1}^n$ is an $n$-bit long moving window over the stream, starting at 
time step $t_1=0$, i.e., $ \{u_{t_1}^n|0\le t_1\le\mathcal{T}-\tau-n\}$. To calculate 
the entropy of the reservoir we consider the collection of instantaneous reservoir 
states $s_{t_2}$ at time step $t_2$, i.e., $\{s_{t_2}|\tau+n\le t_2\le\mathcal{T}\}$. 
The output pattern is calculated using the output of the $\mathcal{B}_n(t)$ task. A 
subtlety arises while calculating the reservoir entropy; since the reservoir follows 
deterministic dynamics, if $L=0$, where the input signals does not perturb the 
reservoir, then the reservoir dynamics will be identical when one repeats the 
experiment. For reservoirs with chaotic dynamics, where each $s_{t_2}$ is 
unique, we have a many-to-one mapping between the reservoir states and the 
output patterns and therefore the reservoir states appear to be capable of 
reconstructing the output completely. To get the correct result, one must calculate 
the entropy over many streams. In this case, since the corresponding output 
patterns change every time, the mapping between the reservoir state and the 
output will not appear predictive of the output.
We feed the reservoir with 50 randomly chosen input sequences of length 50. 
The entropies $\mathcal{H}_I$, $\mathcal{H}_R$, and $\mathcal{H}_O$ are 
calculated using the states the input, the reservoir, and the desired output takes 
during this 50 time interval. Note there is no need to have an output layer in these 
experiments and the calculations are independent of training mechanism. 

Figure~\ref{fig:mutualinfo} illustrates $\mathcal{I}(I:R)$, and $\mathcal{I}(R:O)$ as 
a function of $L$ for reservoirs of $\langle K\rangle \in \{1, 2, 3 \}$. 
For comparison, we have also included $\mathcal{H}_I$, $\mathcal{H}_R$, and $
\mathcal{H}_O$. In an ideal reservoir in which the reservoir contains 
all the information from the input $\mathcal{I}(I:R)=\mathcal{H}_I$, and $
\mathcal{I}(R:O)=\mathcal{H}_O$, indicating that the reservoir contains the 
required information to reconstruct the desired output perfectly. For $\langle K
\rangle=1$ we see that growing $L$ increases $\mathcal{I}(I:R)$ and $\mathcal{I}
(R:O)$ to a maximum level below the ideal values even for $L=500$, where all 
the nodes in the reservoirs are receiving the input. These systems do not have 
enough capacity to calculate the desired output perfectly. For $\langle K
\rangle=2$, we see that mutual information increases and reaches the ideal level 
at $L=20$. In these systems, the sparse connectivity between input and reservoir 
is enough to provide all the required information about the input to the reservoir. 
We see that at the same level of $L$ the reservoir dynamics are completely 
predictive of the output. For $\langle K\rangle=3$, the intrinsic dynamics of the 
reservoir are very rich (supercritical dynamics) and the mutual information 
between the reservoir and output reaches its peak at $L=5$. In these systems a 
small perturbation quickly spreads. The reservoir at this perturbation level will 
have enough information to reconstruct the output.
\begin{figure}[h!]
\centering
\subfigure[]{
\includegraphics[width=1.55in]{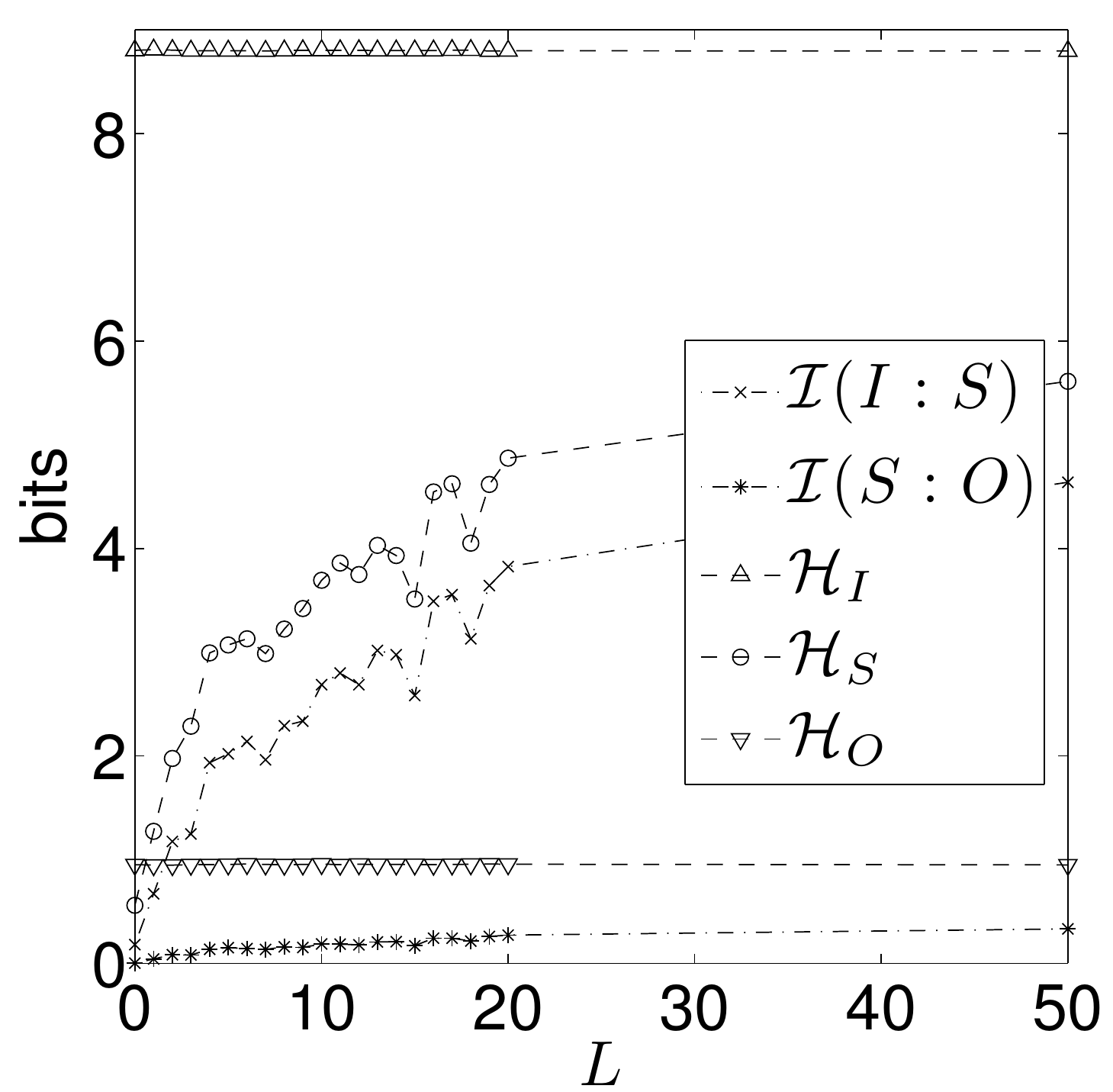}
\label{fig:1maxx20}
}
\subfigure[]{
\includegraphics[width=1.55in]{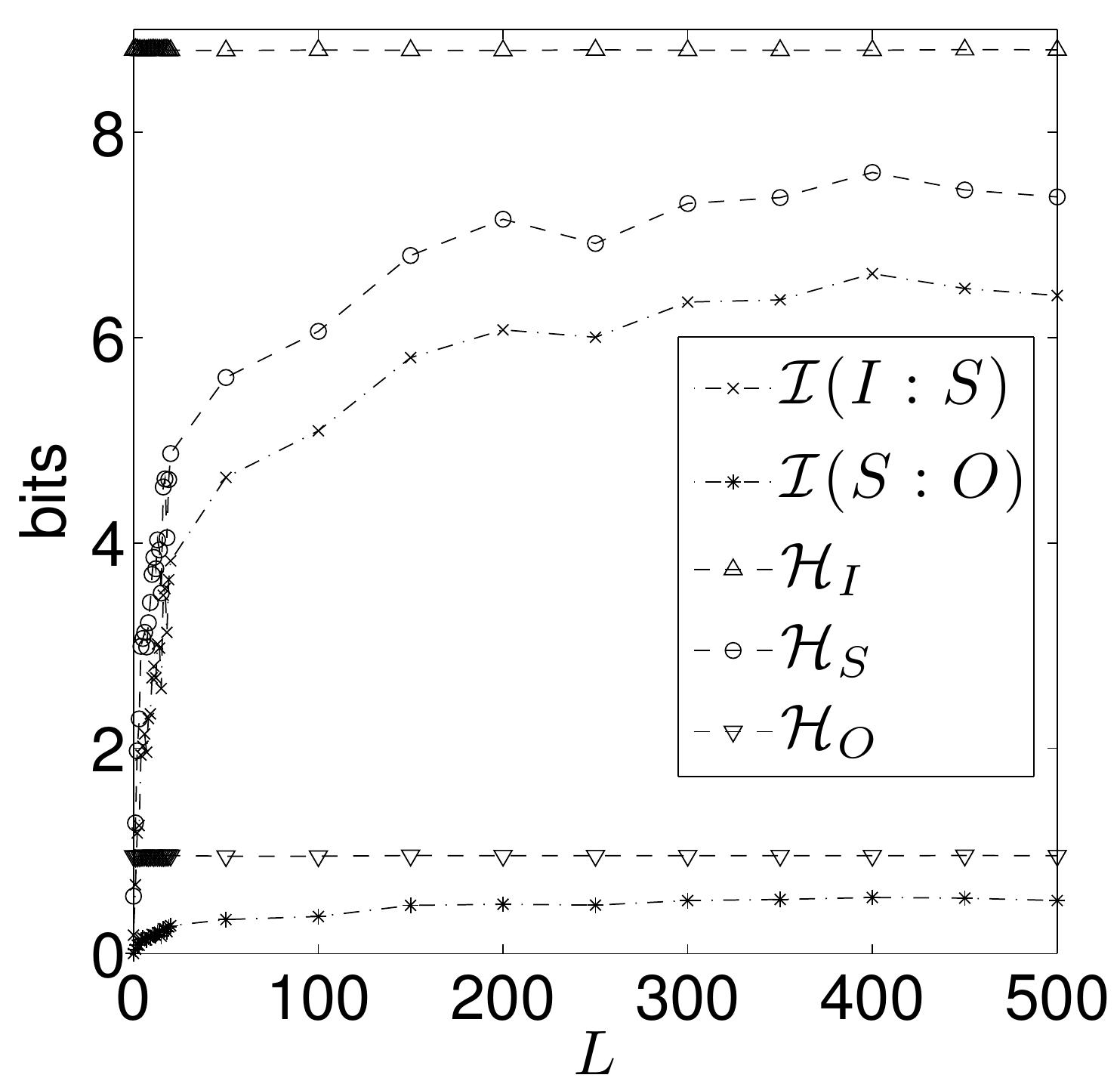}
\label{fig:1maxx500}
}
\subfigure[]{
\includegraphics[width=1.55in]{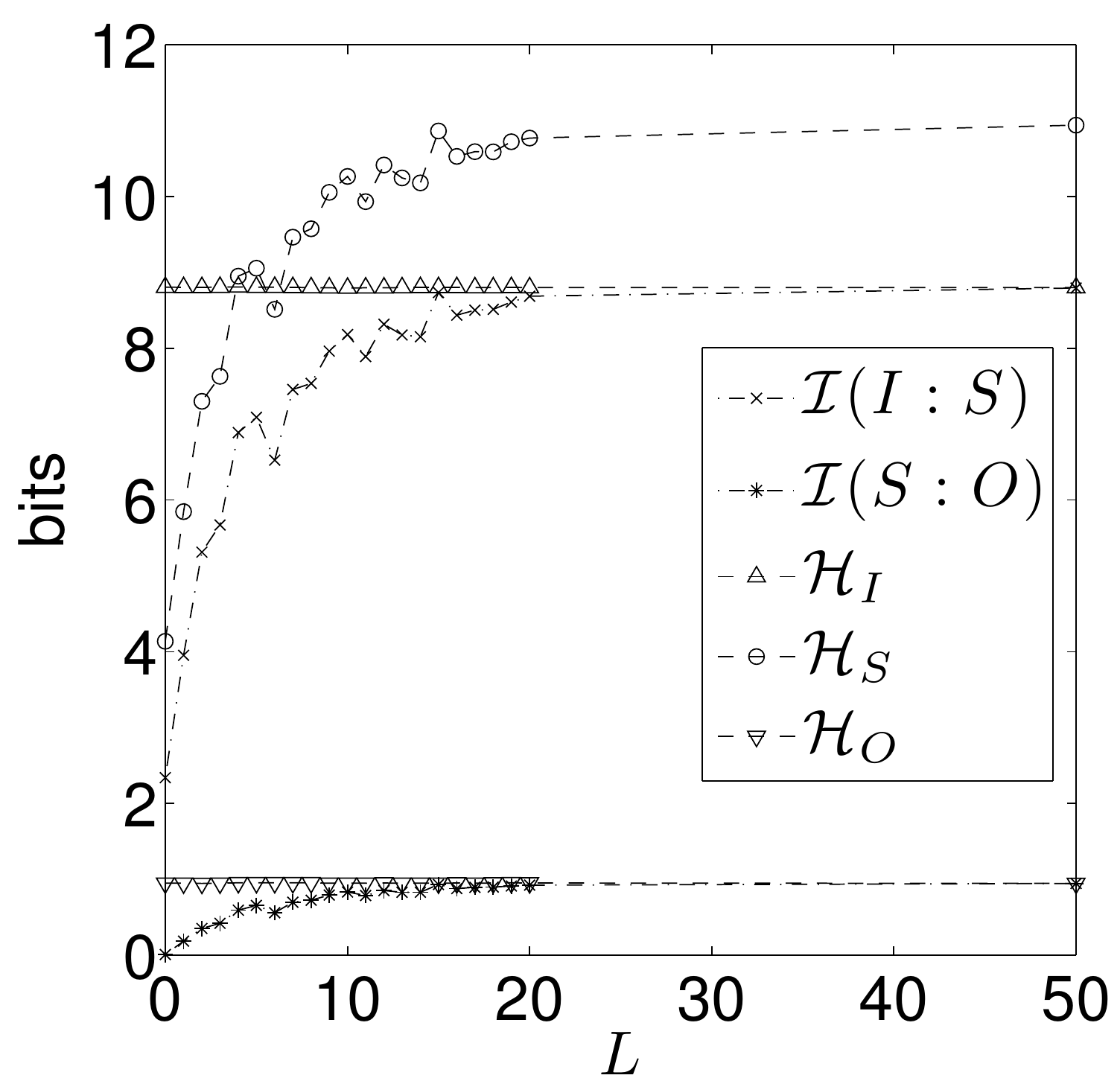}
\label{fig:2maxx20}
}
\subfigure[]{
\includegraphics[width=1.55in]{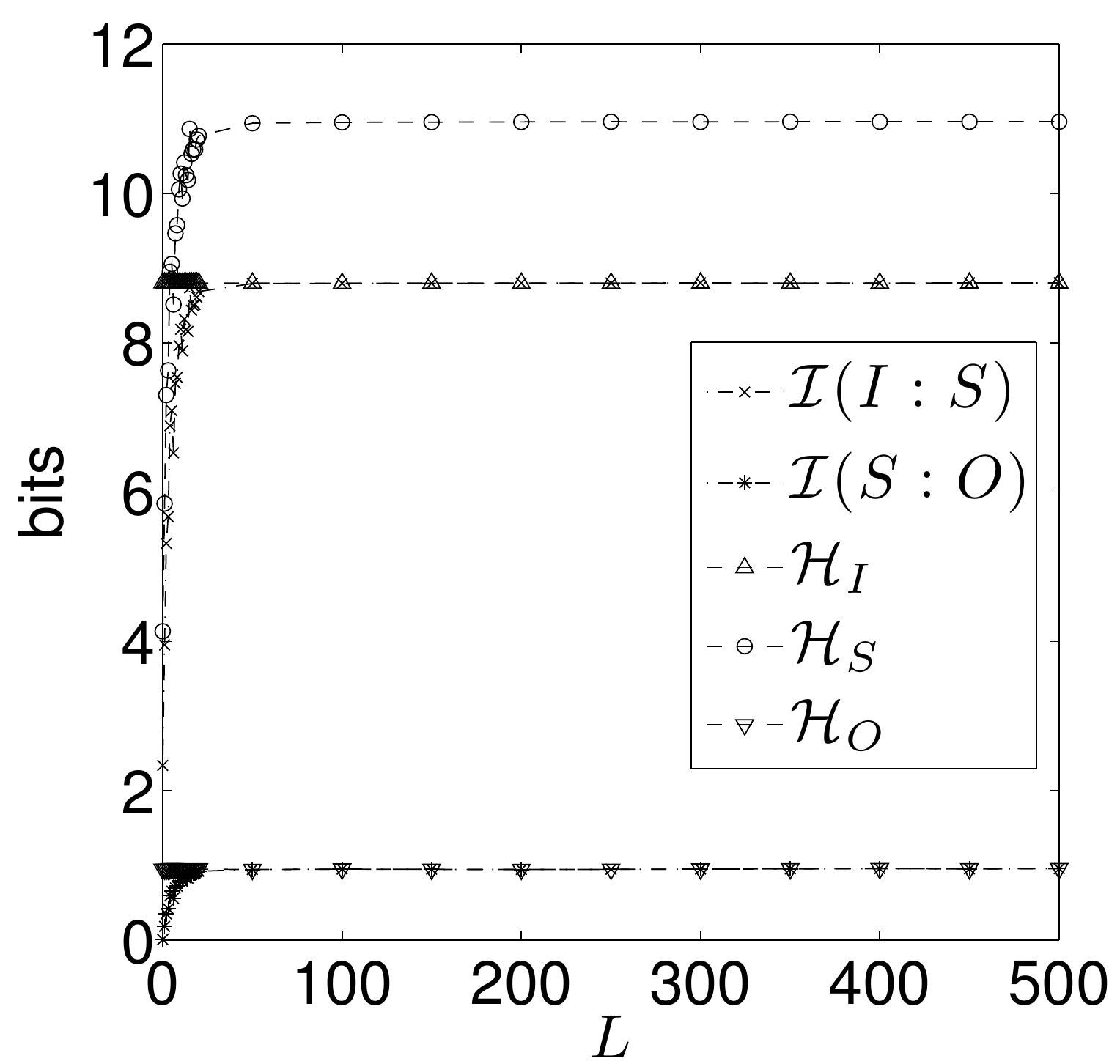}
\label{fig:2maxx500}
}
\subfigure[]{
\includegraphics[width=1.55in]{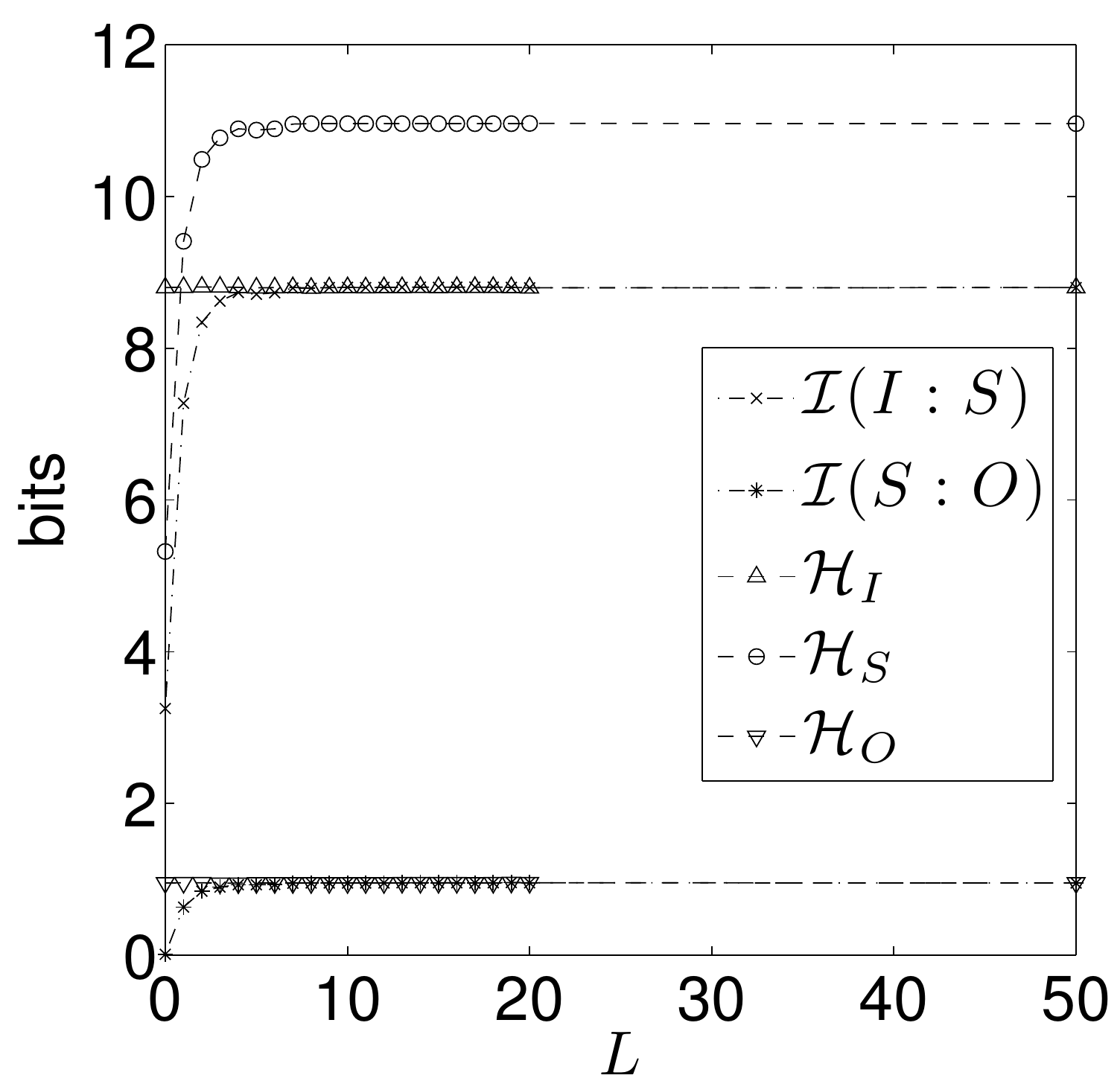}
\label{fig:3maxx20}
}
\subfigure[]{
\includegraphics[width=1.55in]{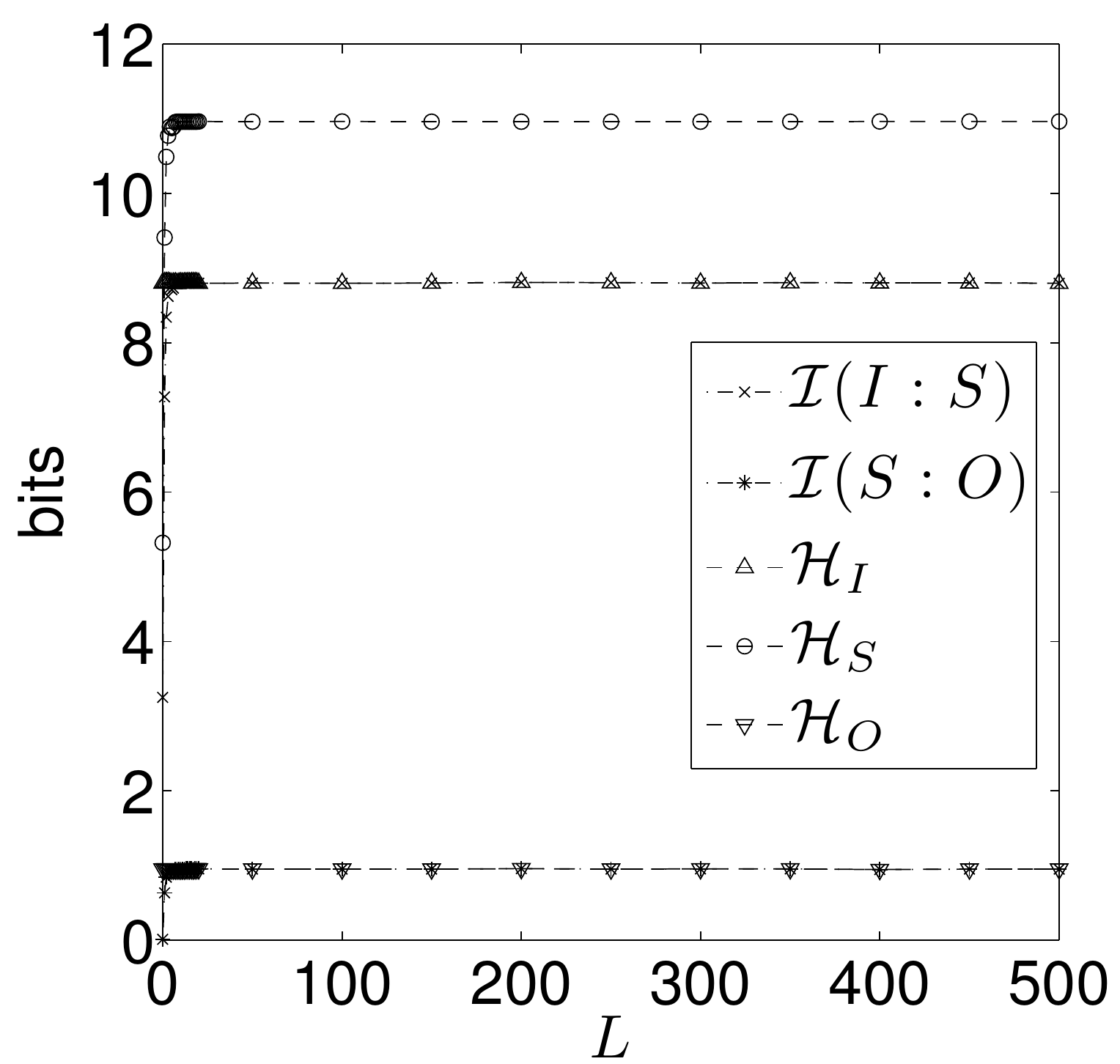}
\label{fig:3maxx500}
}
\caption{Mutual information between the input and reservoir $\mathcal{I}(I:R)$ and reservoir and output $\mathcal{I}(R:O)$. The results are for: $\langle K\rangle=1.0$~\subref{fig:1maxx20} and~\subref{fig:1maxx500}, $\langle K\rangle=2.0$~\subref{fig:2maxx20} and ~\subref{fig:2maxx500}, $\langle K\rangle=3.0$~\subref{fig:3maxx20} and~\subref{fig:3maxx500}. 
We have also included the intrinsic information in the input stream, reservoir, and output. For all of these cases $\tau=1$.  
In an ideal reservoir $\mathcal{I}(I:R)=\mathcal{H}_I$ and $\mathcal{I}(R:O)=\mathcal{H}_O$. For $\langle K\rangle=1.0$, as $L$ 
grows both $\mathcal{I}(I:R)$ and $\mathcal{I}(R:O)$ grow to a maximum level below the ideal level. The reservoir in these systems 
does not carry enough information for the output layer to solve the task. For $\langle K\rangle=2.0$, as $L$ grows, 
the mutual information grows and reaches the sufficient level at $L=20$. For $\langle K\rangle=3.0$, mutual information 
peaks near $L=5$. Small perturbation from the input provide enough information to the reservoir to reconstruct the desired output. }
\label{fig:mutualinfo}
\end{figure}

\subsection{Task Solving}

\begin{figure}[h!]
\centering
\subfigure[]{
\includegraphics[width=1.55in]{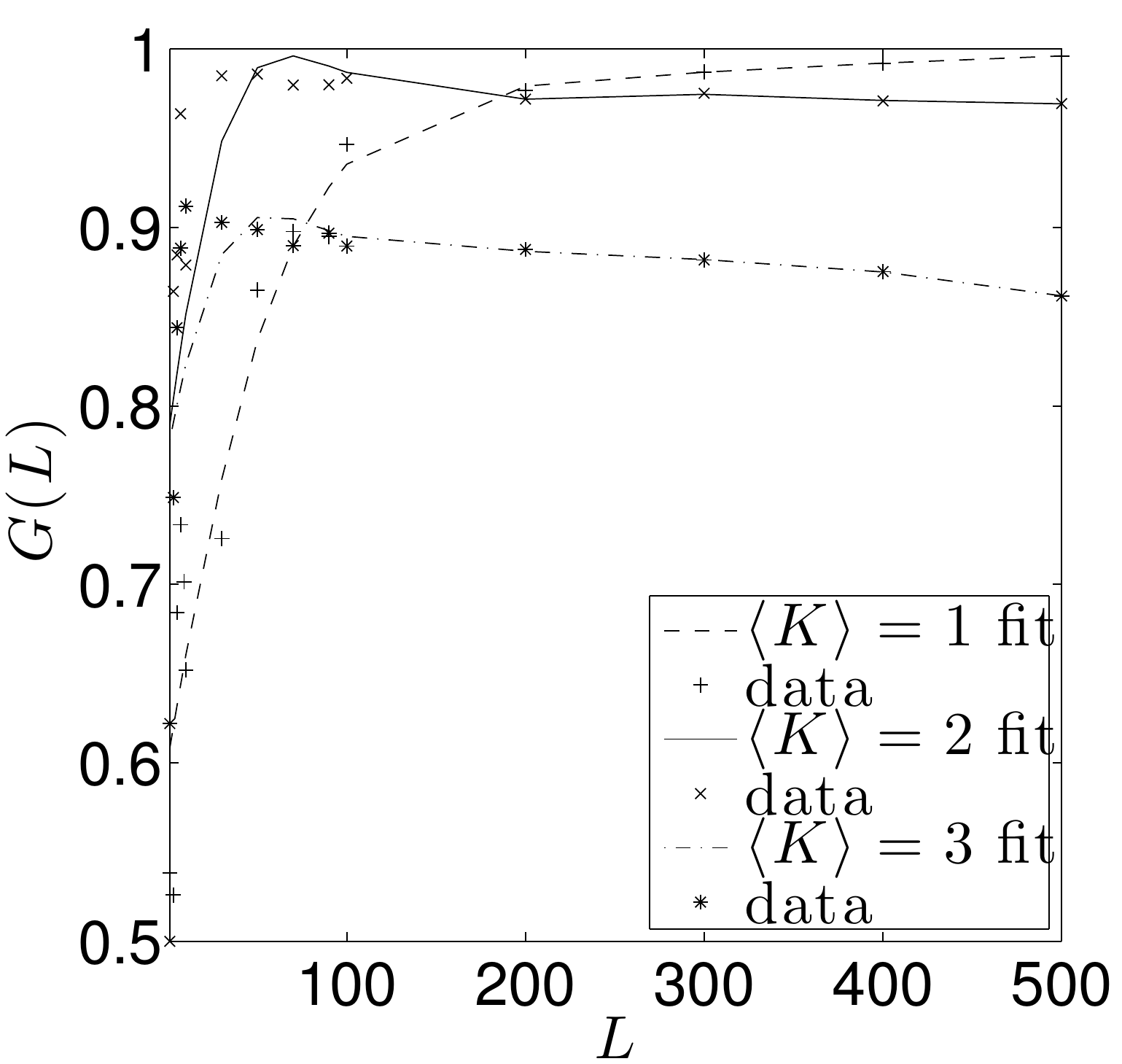}
\label{fig:DNS_n3_tau1_T10}
}
\subfigure[]{
\includegraphics[width=1.55in]{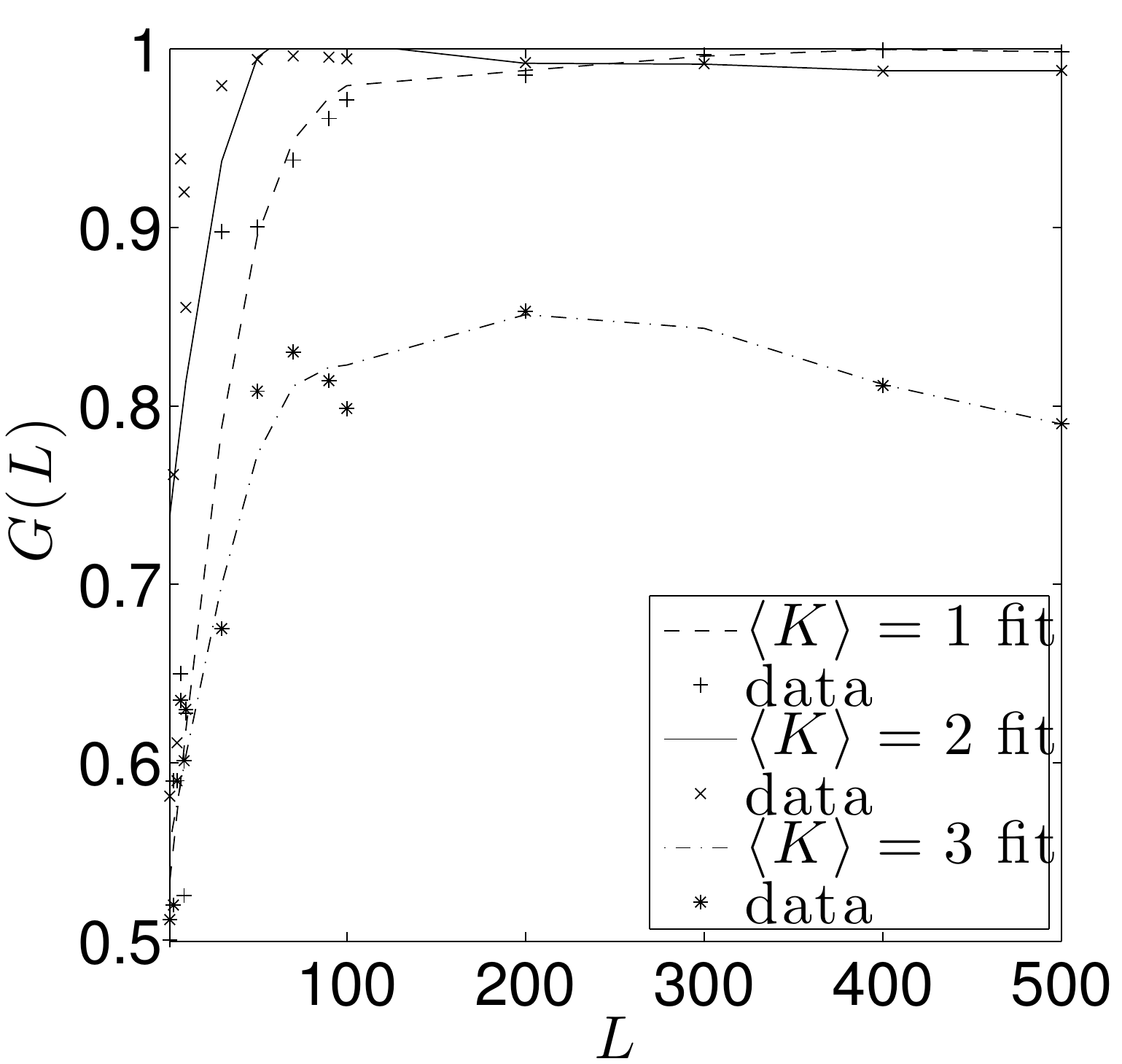}
\label{fig:DNS_n3_tau1_T100}
}
\subfigure[]{
\includegraphics[width=1.55in]{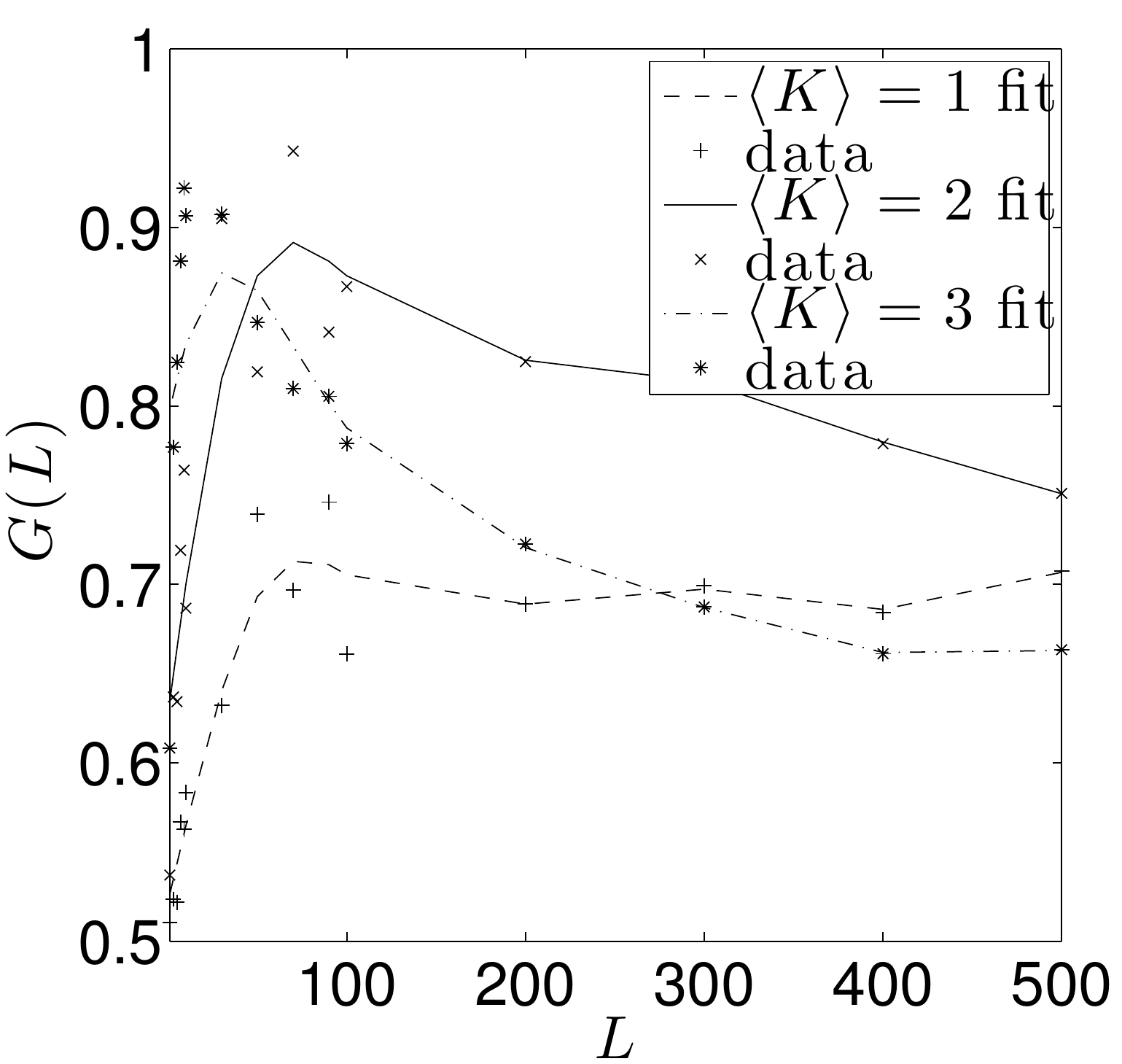}
\label{fig:DNS_n7_tau1_T10}
}
\subfigure[]{
\includegraphics[width=1.55in]{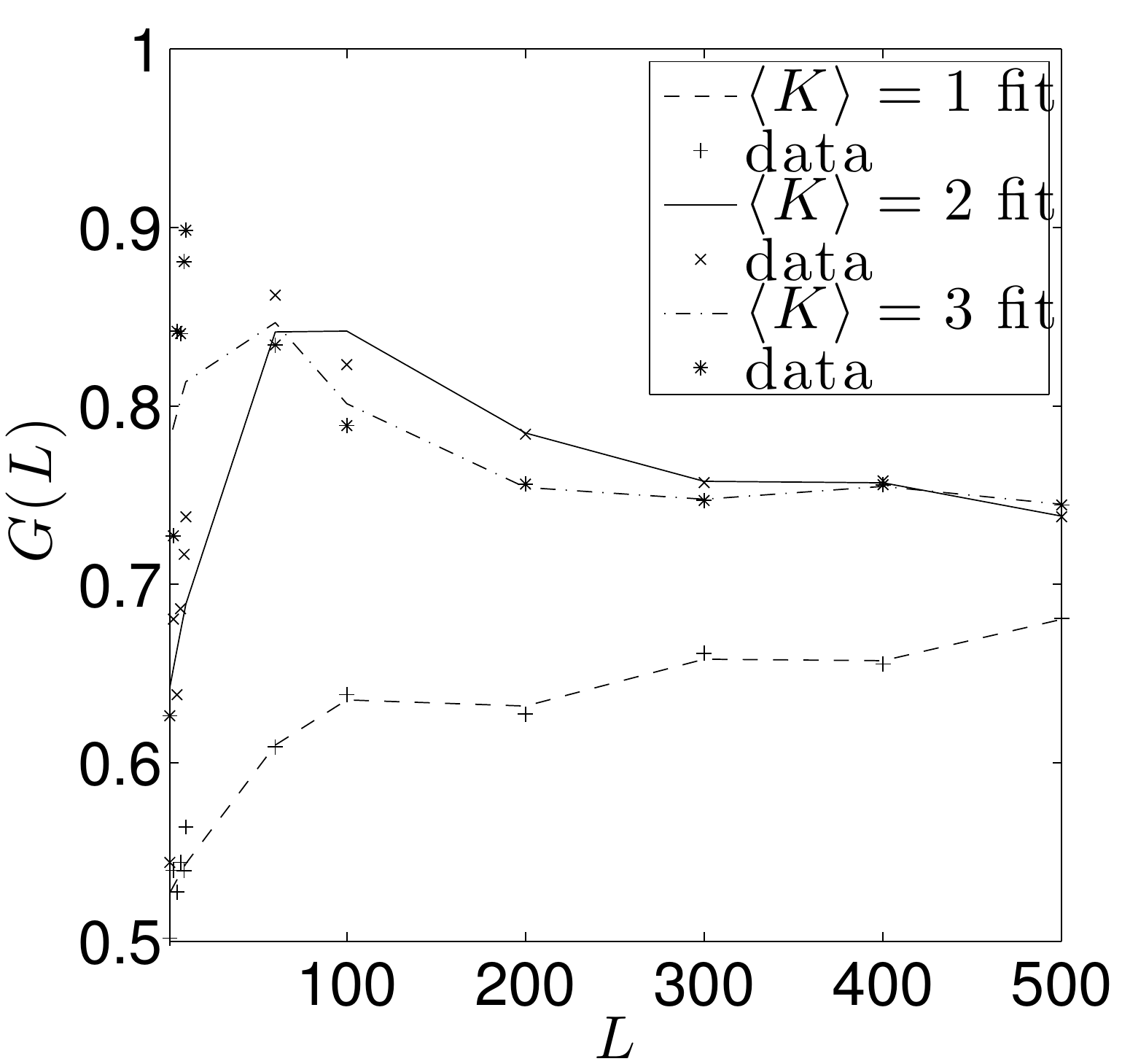}
\label{fig:DNS_n9_tau1_T10}
}
\caption{The generalization capability $G$ of an RBN-RC device $\mathcal{M}$ on the task $\mathcal{B}_{n}$ is dependent on $L$ and $\langle K \rangle$, as well as task parameters $n$ and $\mathcal{T}$. Here the parameters are: $n=3,\mathcal{T}=10$~\subref{fig:DNS_n3_tau1_T10}, $n=3,\mathcal{T}=100$~\subref{fig:DNS_n3_tau1_T100}, $n=7,\mathcal{T}=10$~\subref{fig:DNS_n7_tau1_T10}, $n=9,\mathcal{T}=10$~\subref{fig:DNS_n9_tau1_T10}. Notably, chaotic networks 
achieve their maximum generalization capability with a lower $L$ than ordered networks. Ordered networks possess little memory
and so their performance drops as $n$ increases. On the other hand, chaotic networks perform poorly with $\mathcal{T} = 100$ as opposed to $\mathcal{T} = 10$, due to an inadequately fading
memory.}
\label{fig:DNS}
\end{figure}

\begin{figure}[h!]
\centering
\subfigure[]{
\includegraphics[width=1.55in]{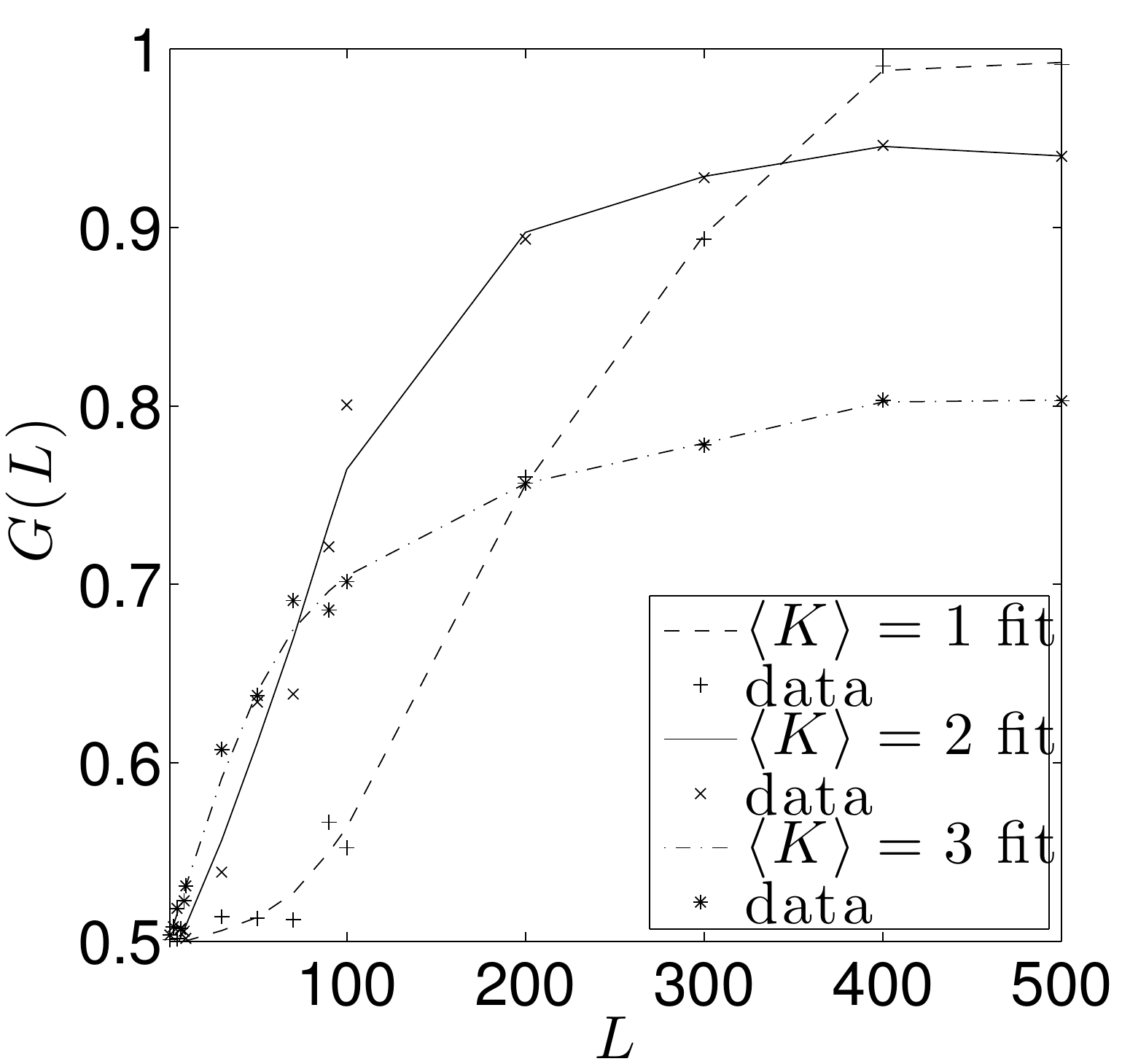}
\label{fig:PAR_n3_tau1_T10}
}
\subfigure[]{
\includegraphics[width=1.55in]{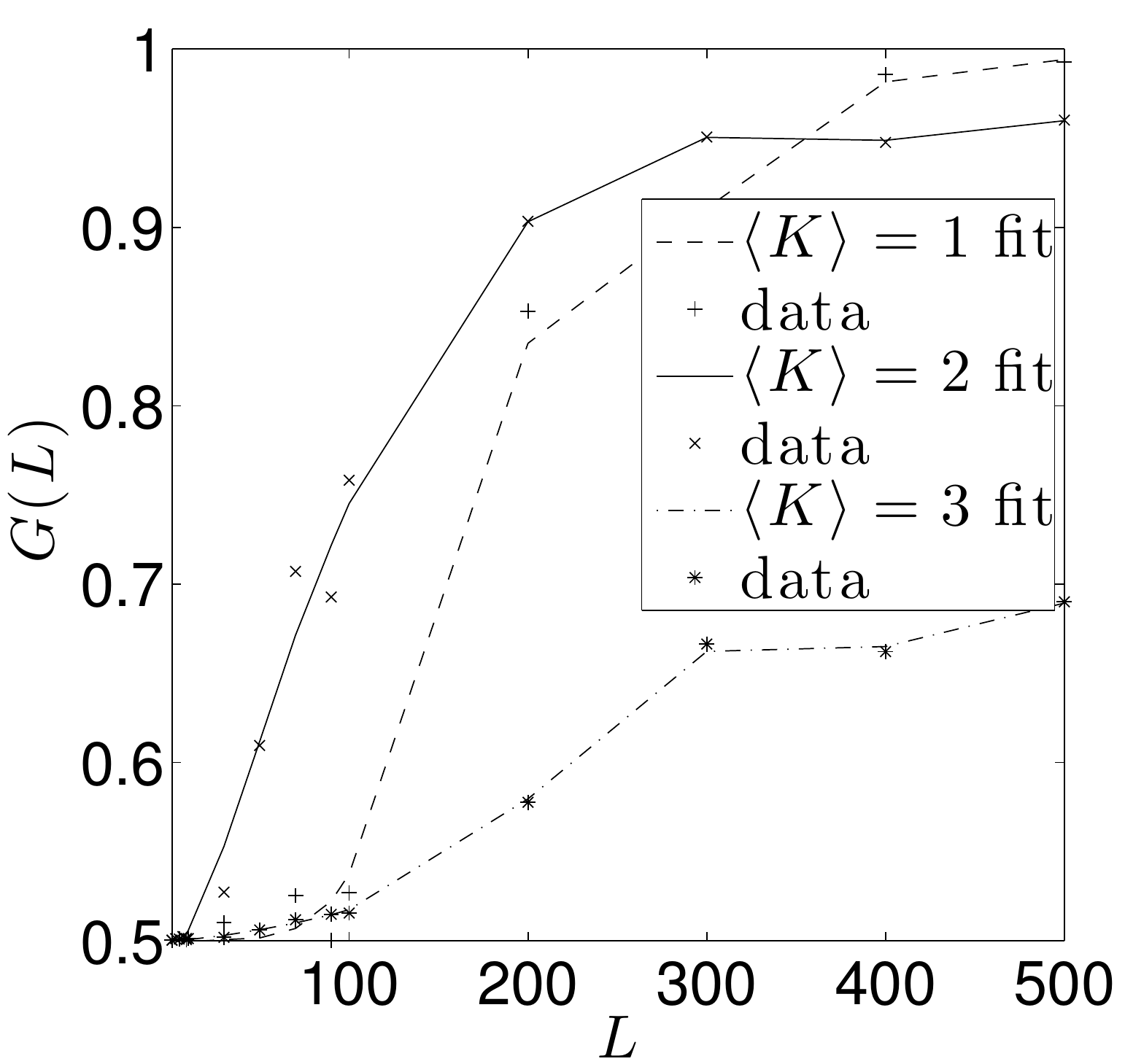}
\label{fig:PAR_n3_tau1_T100}
}
\subfigure[]{
\includegraphics[width=1.55in]{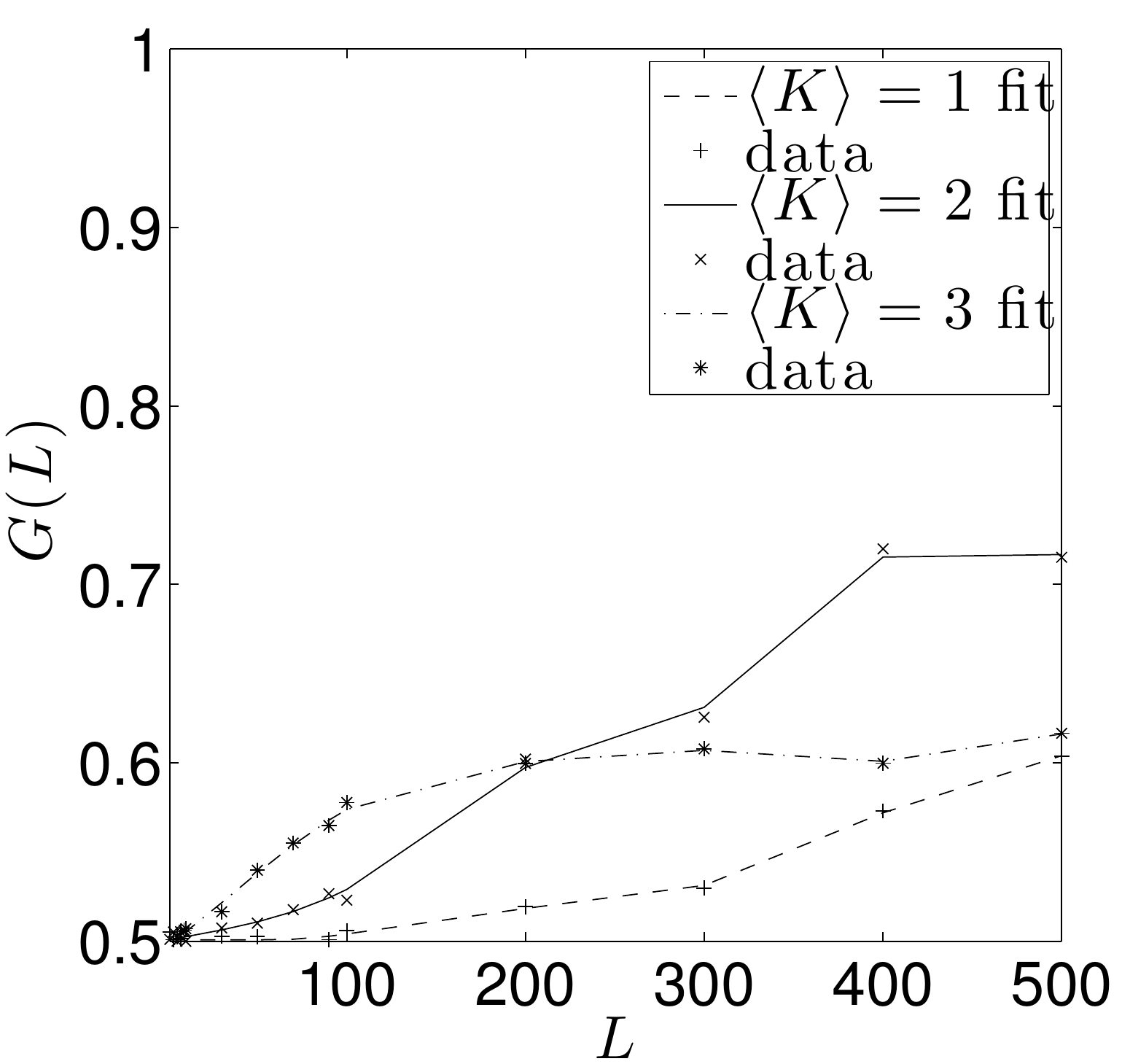}
\label{fig:PAR_n5_tau1_T10}
}
\subfigure[]{
\includegraphics[width=1.55in]{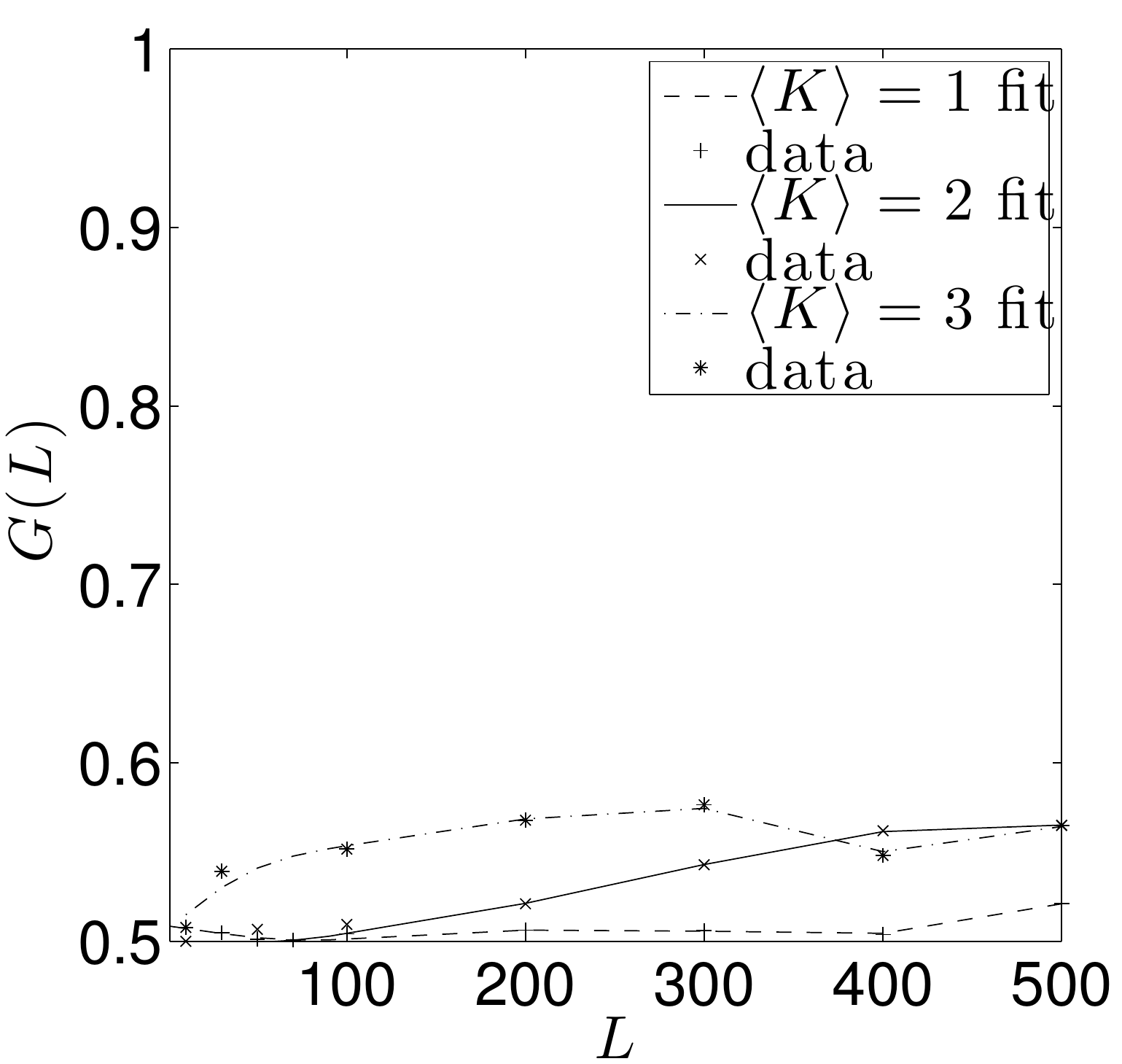}
\label{fig:PAR_n7_tau1_T10}
}
\caption{{The generalization capability $G$ of an RBN-RC device $\mathcal{M}$ on the task $\mathcal{A}_{n}$ is dependent on $L$ and $\langle K \rangle$, and task parameters $n$ and $\mathcal{T}$. Here the parameters are: $n=3,\mathcal{T}=10$~\subref{fig:PAR_n3_tau1_T10}, $n=3,\mathcal{T}=100$~\subref{fig:PAR_n3_tau1_T100}, $n=5,\mathcal{T}=10$~\subref{fig:PAR_n5_tau1_T10}, $n=7,\mathcal{T}=10$~\subref{fig:PAR_n7_tau1_T10}. Due to high sensitivity to the initial perturbations, 
the generalization capability of chaotic networks drop as the length of the input stream increases from $\mathcal{T} = 10$ to $\mathcal{T} = 100$. 
Ordered networks possess little short-term memory and are least robust to an increase in $n$, with generalization capability on $\mathcal{A}_{7}$ no better than chance.}}
\label{fig:PAR}
\end{figure}

We calculate the generalization capability $G$ of RBN-RC devices with $\langle 
K \rangle \in \{1,2,3\}$, $N=500$, and $L \in (0,500]$ on the $\mathcal{B}_{n}$ 
and $\mathcal{A}_{n}$ tasks with random input streams of length $\mathcal{T} \in 
\{10,100\}$. For each set of parameters, we instantiate, train, and test $30$ RC 
devices. Figs.~\ref{fig:DNS} and~\ref{fig:PAR} present cubic spline fits to the 
average of these results. We observe in Figs.~\ref{fig:DNS} and~\ref{fig:PAR} 
that critical dynamics provide the most robust generalization capability in task 
solving. Ordered and chaotic reservoirs can evidently solve tasks under certain 
circumstances. However, the ordered networks are limited by little short-term 
memory, while the chaotic networks accumulate extraneous information from past 
perturbations and demonstrate reduced performance as the length of the input 
stream increases.

\subsubsection{Average Reservoir Indegree $\langle K \rangle = 1$} When $n$ 
of $\mathcal{A}_{n}$ and $\mathcal{B}_{n}$ is small, there is very little memory 
and processing required by the reservoir, and so RC devices in which the 
reservoir has $\langle K \rangle = 1$ can achieve perfect generalization 
$G$ for $\mathcal{B}_{3}$ and $\mathcal{A}_{3}$ 
(cf. Figs.~\ref{fig:DNS_n3_tau1_T10} and~\ref{fig:PAR_n3_tau1_T10}). However, ordered 
networks are dominated by \textit{fading memory}, hence the dynamics do not 
retain enough information about past perturbations to achieve high accuracy 
when $n$ increases. Since the dynamics of ordered networks are only 
determined by their most recent perturbations, the length of the input stream $
\mathcal{T}$ is irrelevant for the task solving capability, which explains why the 
generalization $G$ of ordered networks computing $\mathcal{B}_{3}$ is very 
similar when $\mathcal{T} = 10$ and $\mathcal{T} = 100$, as seen 
in Figs.~\ref{fig:DNS_n3_tau1_T10} and~\ref{fig:DNS_n3_tau1_T100} 
respectively.

Since memory fades quickly in an ordered reservoir, input cannot propagate 
swiftly through the network. Moreover, a $\langle K \rangle = 1$ network will
almost certainly possess islands. These islands will be unreachable by an input 
stream that does not strongly perturb the system. In addition, 
figure~\ref{fig:1maxx500} demonstrates that an ordered network with $\langle K \rangle = 1$ 
increases its mutual information between input and reservoir $\mathcal{I}(I:R)$ as 
$L$ increases from $0$ to $N$. Because of this, we see in Figs.~\ref{fig:DNS} 
and~\ref{fig:PAR} that increasing $L$ tends to result in higher accuracy on $
\mathcal{B}_{n}$ and $\mathcal{A}_{n}$ respectively. Therefore, an increase of 
$L$ can only increase the performance of the reservoir.

\subsubsection{Average Reservoir Indegree $\langle K \rangle = 3$} Chaotic 
reservoirs, represented here by $\langle K \rangle = 3$, are dominated by the 
\textit{separation property}. As a result, the $G$ of the chaotic networks are least 
affected by increasing the window $n$. However, high performance is only 
possible when the input stream is sufficiently small, such as $\mathcal{T} = 10$. 
In Figs.~\ref{fig:DNS_n9_tau1_T10} and~\ref{fig:PAR_n3_tau1_T10} the 
stream length is $\mathcal{T} = 10$ and the $G$ of the chaotic network is high. 
However, when the length of the input stream is increased to $\mathcal{T} = 
100$, the performance drops significantly, even while the performance of 
networks with lower connectivities remain relatively unchanged (cf. Figs.~\ref{fig:DNS_n3_tau1_T10} and~\ref{fig:DNS_n3_tau1_T100}).
Though no longer relevant to the RC device, early perturbations have a 
significant effect on the dynamics, which makes it difficult for the output layer to 
extract information about the more recent input. On the other hand, if the input 
stream is sufficiently short, chaotic reservoirs have less time to be distorted by 
early input.

In a network which is chaotic due to high connectivity, there will be fewer and 
larger connected components than in those which are less densely connected~\citep{Erdos:1959p1849}. Therefore, the minimal $L$ needed to adequately 
distribute the input signal in a $\langle K \rangle = 3$ network is less than the 
other connectivities explored here (cf. Fig.~\ref{fig:3maxx500}). Also, a 
chaotic reservoir can effectively increase its computational capability as predicted 
by $\Delta$ by reducing its $L$, which increases the time it takes for 
perturbations to spread through the system (cf. Fig.~\ref{fig:K3dot0T103D}). 
Evidently, the chaotic system uses this strategy in computing $\mathcal{B}_{n}$ 
when $\mathcal{T} = 10$, as seen in figure~\ref{fig:DNS}. However, this behavior 
is not observed for the highly non-linear parity task $\mathcal{A}_{n}$. We 
speculate that, due to the complexity of the task, \textit{separation capability} is 
more significant than it is in $\mathcal{B}_{n}$; this causes a strategy which 
maximizes the \textit{separation property} by increasing $L$ to be optimal.

\subsubsection{Average Reservoir Indegree $\langle K \rangle = 2$} As 
observed in figure~\ref{fig:delta_K_tau}, the difference between the 
\textit{separation property} and \textit{fading memory} tends to be maximized with 
near-critical connectivity $\langle K \rangle = 2$. This is evident in our task 
solving results: when the $n$ of $\mathcal{B}_{n}$ and $\mathcal{A}_{n}$ 
increases, these systems do not show the dramatic drop in $G$ 
that the ordered systems do (cf. Figs.~\ref{fig:DNS} and~\ref{fig:PAR}). 
Simultaneously, the $G$ of these systems are unaffected by an increase in the 
stream length $\mathcal{T}$, in contrast to chaotic networks. In 
figure~\ref{fig:2maxx500} we observe that with $L < 20$ the input signal cannot adequately 
propagate the input signal, which is demonstrated by a lower $G$ for very small 
$L$ in Figs.~\ref{fig:DNS} and~\ref{fig:PAR}. However, increasing $L$ in task 
solving appears to afford more of a benefit than simply increasing the information 
about the input stream. In both Figs.~\ref{fig:DNS} and~\ref{fig:PAR} we see that 
the best $G$ of the critical networks occurs after the system has already achieved 
maximal $\mathcal{I}(I:R)$ between input stream and reservoir dynamics.

\section{Summary and Discussion}

We investigated the computational capabilities of random Boolean
networks when used as the dynamical component in reservoir computing
devices.  We found that computation tends to be maximized at the
critical connectivity $\langle K_{c} \rangle = 2$.  However, in RC,
the reservoir is continuously perturbed, and both the size of the
perturbations as well as the length of time that the reservoir is
perturbed for must be taken into account, along with the chaoticity of
the dynamics.  If the input stream is sufficiently short, then chaotic
systems can still perform quite well, but as the length of the input
stream increases, these networks can no longer differentiate and
generalize on subsets of the input stream, as the past perturbations,
which may no longer be relevant to the computation, are dominating the
dynamics.  On the other hand, ordered networks can perform well,
independent of the length of the input stream, as long as the window
of computation is sufficiently small, as an ordered system retains
little information about perturbations in the past.

A network view of the RC device can also give us more insight as to
why connectivity influences performance.  If the reservoir acts on the
input stream as a set of spatiotemporal kernels, a suitable reservoir
needs to include a diverse set of kernels.  In
\cite{PhysRevLett.108.128702}, we saw that at the connectivity
$\langle K_{c} \rangle = 2$ the network shows maximal topological
diversity and dynamics.  A reservoir with connectivity $\langle K_{c}\rangle = 2$ therefore can act as many networks of the same
connectivity, each acting as different kernel.

In \cite{natschlager2004} and \cite{4905041020100501} it was shown
that optimal computation occurs in recurrent neural networks at the
critical points, and our results provide an additional example of
this, in a binary, heterogeneous reservoir.  In RC, we continuously
perturb the reservoir and so the underlying RBN of our model is not a
closed system. Therefore, computation cannot be dependent on
attractors and must be enabled by the dynamics of the RBN. However,
in some circumstances the network dynamics can fall into an attractor
temporarily or indefinitely, due to frozen dynamics, inadequate
distribution of the input signal, or a non-random input stream. Therefore, RC is a novel framework in which to explore the capacity of
RBN dynamics for information processing.  RBNs have been studied under
other task solving scenarios; in \citet{PhysRevLett.108.128702}
networks evolve towards criticality, although computation is still
performed by attractors.  Our study shows that unlike the findings in
\cite{mitchell93}, for RBNs there is a strong connection between
computation and dynamics, and optimality of the computation is
evidently due to critical dynamics in the network. Despite the
differences between externally perturbed RBNs in RC and RBNs explored
as a closed system, we nevertheless observe that critical RBNs are
indeed optimal for reservoir computing. {Criticality
  also plays an important role in biological systems that often
  require an optimal balance between stability and adaptability. For
  example, it has been shown by using compelling theoretical and
  experimental evidence that gene regulatory networks---which are
  commonly modeled by RBNs---are indeed critical
  \cite{Ramo:2006,Nyker:2008}.} 

Our conclusion provides an intriguing link between disparate
usages of RBN. By providing evidence that critical dynamics are
desirable for heterogeneous substrates in RC, our findings may be
relevant to the development of devices which exploit the intrinsic
information processing capabilities of heterogeneous, physical systems
such as biomolecular or nanoscale device networks.

\section{Acknowledgments}
This work was supported by NSF grants No. 1028120 and No. 1028378 as
well as Portland State University's Maseeh College of Engineering \&
Computer Science Undergraduate Research \& Mentoring Program.


\vspace{-5mm}

\begin{thebibliography}{99}
\bibitem[Lukosevicius and Jaeger, 2009]{Lukosevicius:2009p1443}
M. Lukosevicius and H. Jaeger, Comput. Sci. Rev. {\bf 3}, 127--149 (2009).

\bibitem[Fernando and Sojakka, 2003]{Fernando2003}
C. Fernando and S. Sojakka, ECAL 2003, (Springer-Verlag, Berlin, 2003), LNAI 2801, p. 588--597.

\bibitem[Haelman, 2010]{Haselman:2010p657}
M. Haselman and S. Hauck, Proc. IEEE, {\bf 98}, 11--38, (2010).

\bibitem[Maass et~al., 2002]{Maass:2002p1444}
W. Maass, T. Natschl{\"a}ger, and H.  Markram, Neural Comput., {\bf 14}, 2531--60, (2002).

\bibitem[Jaeger, 2001]{Jaeger:2001p1442}
H. Jaeger, St. Augustin: German National Research Center for Information Technology Technical Report GMD Rep. 148, (2001).
 
\bibitem[Kauffman, 1969]{Kauffman:1969p2786}
S.~A.\ Kauffman, J.\ Theor.\ Biol.\ {\bf 22}, 437 (1969)

\bibitem[Rohlf et~al., 2007]{rohlf07_prl}
T. Rohlf, N. Gulbahce, and C. Teuscher, Phys. Rev. Lett., {\bf 99}, 248701 (2007).

\bibitem[Packard, 1988]{packard1988}
N. H. Packard, in Dynamic Patterns in Complex Systems, ed. by J. A. S. Kelso, A. J. Mandell, and M. F. Shlesinger (World Scientific, Singapore, 1988), p. 293-301.

\bibitem[Mitchell et~al., 1993]{mitchell93}
M. Mitchell, J. P. Crutchfield, and P. T. Hraber, Complex Syst., {\bf 7}, 89--130 (1993).

\bibitem[Goudarzi et~al., 2012]{PhysRevLett.108.128702}
A. Goudarzi, C. Teuscher, N. Gulbahce, and T. Rohlf, Phys. Rev. Lett. {\bf108}, 128702 (2012).

\bibitem[Snyder et~al., 2012]{snyder2012}
D. Snyder, A. Goudarzi, and C. Teuscher, in Proc. Thirteenth Int'l Conference on the Simulation and Synthesis of Living Systems (ALife 13), (MIT Press, Cambridge, 2012), p. 259--266.
  
\bibitem[Natschl{\"a}ger et~al., 2004]{natschlager2004} 
T. Natschl{\"a}ger, N. Bertschinger, and R. Legenstein, in Advances in Neural Information Processing Systems, edited by L. K. Saul, Y. Weiss, and L. Bottou (MIT Press, Cambridge, 2004), {\bf17}, p. 147--152.

\bibitem[B{\"u}sing et~al., 2010]{4905041020100501}
L. B{\"u}sing, B. Schrauwen, and R. Legenstein, Neural Comput.,  {\bf 22} 1272--1311, (2010).

\bibitem[Erd{\"o}s and R{\'e}nyi, 1959]{Erdos:1959p1849}
P. Erd{\"o}s and A. R{\'e}nyi
\newblock {\em Publ. Math. Debrecen}, {\bf6}, 290--297, (1959).

\bibitem[Derrida et~al., 1986]{Derrida:1986}
B. Derrida and Y. Pomeau, Europhys. Lett. 1, 45 (1986).

\bibitem[Shannon, C. E., 1948]{Shannon:1948}
C. E. Shannon,  Bell Sys. Tech. J., {\bf 27} 379--423, (1948).

\bibitem[P. Ramo et~al., 2006]{Ramo:2006}
P. R\"am\"o, J. Kesseli, and O. Yli-Harja, J.\ Theor.\ Biol., {\bf 242}, 164--170 (2006).

\bibitem[Nyker, M., 2008]{Nyker:2008}
M. Nyker, N. D. Price, M Aldana, S. A. Ramsey, S. A. Kauffman, L. E. Hood, O. Yli-Harja, and I. Shmulevich, Proc. Natl. Acad. Sci. (USA), {\bf 105}, 1897--1900 (2008).


\end{thebibliography}
\end{document}